\documentclass[aps,prd,longbibliography,onecolumn,showpacs,amsmath,amssymb,superscriptaddress,nofootinbib,floatfix,preprintnumbers,secnumarabic,11pt]{revtex4-1}

\usepackage{graphicx}
\usepackage[dvipsnames]{xcolor}
\newcommand{\remove}[1]{}
\usepackage{xspace}
\definecolor{darkblue}{rgb}{0,0,0.5}
\definecolor{darkgreen}{rgb}{0.1,0,0.3}
\definecolor{darkred}{rgb}{0.6,0,0}
\usepackage[bookmarks=false, bookmarksnumbered=true, colorlinks=true, urlcolor=darkblue, citecolor=darkgreen, linkcolor=darkred, breaklinks=true]{hyperref} 
\usepackage{siunitx}
\usepackage[normalem]{ulem}
\usepackage[capitalise]{cleveref}
\usepackage{multirow}

\usepackage{setspace}

\newcommand{\mueff}{\mu^*}

\newcommand{\lmultau}{$U(1)_{L_{\mu}-L_{\tau}}$\xspace}

\newcommand{\PRDnote}[1]{\textcolor{black}{{#1}}}

\usepackage{fullpage}
\usepackage{geometry}   
\begin{document}
\setstretch{1}

\preprint{IFT-UAM/CSIC-22-130}
\preprint{FTUAM-22-2}

\vspace*{0.7cm}

\title{
Constraints from the duration of supernova neutrino burst on on-shell light gauge boson production by neutrinos
}

\author{David Cerde\~no}
\email{davidg.cerdeno@uam.es}
\affiliation{Instituto de F\' isica Te\'orica, IFT-UAM/CSIC, 28049 Madrid, Spain}
\affiliation{Departamento de F\' isica Te\'orica, Universidad Aut\'onoma de Madrid, 28049 Madrid, Spain}

\author{Marina Cerme\~no}
\email{marina.cermeno@ift.csic.es}
\affiliation{Instituto de F\' isica Te\'orica, IFT-UAM/CSIC, 28049 Madrid, Spain}
\affiliation{Centre for Cosmology, Particle Physics and Phenomenology (CP3),Universit\'e catholique de Louvain, Chemin du Cyclotron 2,B-1348 Louvain-la-Neuve, Belgium} 

\author{Yasaman Farzan}
\email{yasaman@theory.ipm.ac.ir}
\affiliation{School of Physics, Institute for Research in Fundamental Sciences (IPM), P.O. Box 19395-5531, Tehran, Iran}

\begin{abstract}
In this article, we study the on-shell production of low-mass vector mediators from neutrino-antineutrino coalescence in the core of proto-neutron stars. Taking into account the radial dependence of the density, energy, and temperature inside the proto-neutron star, we compute the neutrino-antineutrino interaction rate in the star interior in the well-motivated \lmultau model. First, we determine the values of the coupling above which neutrino-antineutrino interactions dominate over the Standard Model neutrino-nucleon scattering. We argue that, although in this regime a redistribution of the neutrino energies might take place, making low-energy neutrinos more trapped, this only affects a small part of the neutrino population and it cannot be constrained with the SN~1987A data. Thus, contrary to previous claims, the region of the parameter space where the \lmultau model explains the discrepancy in the muon anomalous magnetic moment is not ruled out. We then focus on small gauge couplings, where the decay length of the new gauge boson is larger than the neutrino-nucleon mean free path, but still smaller than the size of proto-neutron star. We show that in this regime, the on-shell production of a long-lived $Z'$ and its subsequent decay into neutrinos can significantly reduce the duration of the neutrino burst, probing values of the coupling below ${\cal O}(10^{-7})$ for mediator masses between 10 and 100~MeV. This
\PRDnote{disfavours} new areas of the parameter space of the \lmultau model.

\end{abstract}
\maketitle

\section{Introduction}
\label{sec:Intro}

Core-collapse supernovae (SN) are the violent explosions that result from the rapid collapse of giant stars at the end of their thermonuclear evolution. These very energetic phenomena provide a unique window to test new physics beyond the Standard Model (SM), especially in the neutrino sector. Neutrinos are copiously produced during the SN collapse and the subsequent cooling of the resulting proto-neutron star (proto-NS), as confirmed by the observation of the burst from SN~1987A.

The production of new exotic particles inside SN cores is severely constrained by the measured neutrino flux from SN~1987A. For example, if such particles are long-lived and very weakly-interacting, they may efficiently transfer energy from the core and thus induce energy loss that exceeds the measured flux of $3 \times 10^{53}$~erg \cite{Sato:1987rd, Spergel:1987ch,Bahcall:1987ua,Burrows:1987zz,Burrows:1988ah, Choi:1989hi, Raffelt:1990yz, Burrows:1990pk,Schramm:1990pf,Loredo:2001rx,Frieman:1987as,Kolb:1988pe,Berezhiani:1989za,Farzan:2002wx,Heurtier:2016otg, Farzan:2018gtr, Escudero:2019gzq, Suliga:2020jfa, Croon:2020lrf, Shin:2021bvz}. Moreover, if this particle mediates new neutrino interactions with matter fields, the neutrino mean free path can be reduced, increasing the observed $\sim$ $10$~s duration of the neutrino burst from SN~1987A. This has been used to derive bounds on the neutrino scattering cross section with nucleons, which can ultimately be translated into constraints on the parameters of the underlying theories. If the new physics is due to a light mediator between neutrinos and SM particles, the constraints on the mediator mass and couplings are complementary to those obtained from other experimental limits in the range of masses between 1~MeV and 1~GeV, see Refs. 
\cite{Farzan:2018gtr, Suliga:2020jfa, Cerdeno:2021cdz}.

If the mass of the mediator (for example, a new vector $Z^\prime$) is comparable to the temperature inside the proto-NS (MeV scale), the process $\nu \bar\nu \to Z^{\prime*}\to\nu \bar\nu$ can be enhanced when the mediator goes on-shell. These interactions are especially important for the muon and tau neutrinos, for which scatterings with their antiparticles are more frequent due to their vanishing chemical potential. In Ref.~\cite{Kamada:2015era}, it was argued that the on-shell $Z'$ production from neutrino-antineutrino interactions in the SN core can significantly increase  the neutrino diffusion time in the \lmultau model, which poses a challenge to the solution to the muon anomalous magnetic moment, $(g-2)_\mu$. It should be noted that although the SN~1987A neutrino signal was only detected through the positrons produced by electron antineutrinos, thanks to the oscillation phenomenon, neutrinos or antineutrinos of any flavour generated in the SN have a ${\cal O}(1)$ probability to be detected as $\nu_e$ or $\bar{\nu}_e$ after they exit the star core and propagate to Earth. Thus, if muon and tau neutrinos have a diffusion time significantly different from $\sim 1$~s (a factor 10 enhancement is expected for the signal time since the stored thermal energy, $\mathcal{E}_{th}^{tot}$, in matter continues to be emitted in neutrinos even after the first neutrinos escape), this should have been observed as well in the electron neutrinos.

We however argue that because of the momentum conservation in scatterings, the mere short-ranged interaction between neutrinos and antineutrinos inside the proto-NS cannot change the energy flux of neutrino plus antineutrino gas so the neutrino burst duration cannot be affected by neutrino-antineutrino interactions as long as they are short-ranged. This argument is in agreement with that of Ref.~\cite{Dicus:1988jh}, which prevented early bounds on the cross section for neutrino self-interactions applying to very large values, $\sigma_{\nu\nu}\sim 10^{-35}$~cm$^2$ \cite{Manohar:1987ec}, as well as the range of couplings in Ref.~\cite{Kamada:2015era}. It would therefore seem that the solution to the $(g-2)_\mu$ is safe.

In this article, we study certain effects on the SN neutrino signal duration caused by the neutrino-antineutrino coalescence that have been largely overlooked in the literature. We first point out that the Breit-Wigner resonant enhancement of neutrino-antineutrino interactions via MeV scale mediators can lead to a redistribution of the neutrino energies in the SN interior. This can make low-energy neutrinos more trapped, however, this effect depends crucially on the fraction of low-energy neutrinos, which we show is rather small. Then, we focus on the regime where the decay length of the mediator is larger than the neutrino-nucleon mean free path. It has already been demonstrated that if the mediator decays outside the proto-NS, the duration of the neutrino signal can be shortened~\cite{Burrows:1988ah, Choi:1989hi, Raffelt:1990yz, Burrows:1990pk,Schramm:1990pf,Loredo:2001rx,Frieman:1987as,Kolb:1988pe,Berezhiani:1989za,Farzan:2002wx,Heurtier:2016otg, Farzan:2018gtr, Escudero:2019gzq, Suliga:2020jfa, Croon:2020lrf, Shin:2021bvz}. The reason is that neutrino and antineutrino pairs inside the proto-NS core can convert to a $Z'$ within time scales much shorter than the diffusion time and can then be transferred outside the neutrinosphere (the radius at which neutrinos decouple from stellar matter). We extend this analysis to the range where the mediator decays inside the star and demonstrate that the neutrino burst duration can still be significantly reduced. \PRDnote{This allows us to identify regions of the parameter space below couplings of ${\cal O}(10^{-7})$ that are disfavoured by the SN 1987A neutrino burst duration measurement.}

Recent works have also considered some neutrino self-interaction effects on SN dynamics. For example, in Ref.~\cite{Chang:2022aas} strong neutrino self-interactions outside the proto-NS have been used to extract bounds on light scalars based on the SN diffusion time. Likewise, in Ref.~\cite{Shalgar:2019rqe} self-interactions were considered between SN neutrinos and the cosmic neutrino background in order to obtain constraints based on the delay of SN neutrinos to reach the Earth. Although 2 to 4 processes were shown to affect the shock revival in the SN neutrino core, the $\nu\bar\nu\to\nu\bar\nu$ processes were not included, since they do not change the number of neutrinos. 
In addition, Ref.~\cite{Croon:2020lrf} derives cooling bounds by studying neutrino-antineutrino coalescence as well as muon semi-Compton scattering that produce long lived vector bosons which decay outside the star.

The on-shell production of the mediator via neutrino-antineutrino coalescence can also alter the energy spectrum of neutrinos measured at Earth. When the $Z'$ can directly come out of the inner hot core (without being cooled in the outer shells), neutrinos leaving the star will be hotter than in the SM prediction. This has been used in Refs.~\cite{Akita:2022etk,Fiorillo:2022cdq} to constrain the parameter space of models featuring a light scalar boson coupled to neutrinos. However, for the range of couplings, $\sim 6 \times 10^{-8}$, and mediator masses, $ 10-100$~MeV, that we \PRDnote{show to be in tension with the SN 1987A measurements} in this article, the neutrino energy spectrum on Earth will not be distinguishable from the SM prediction. The time duration argument that we study in this article provides a better route to test this region of the parameter range.

In this article, based on the modification of the neutrino flux duration, we re-evaluate previous constraints for MeV scale mediators. As a concrete example, we \PRDnote{find new disfavoured regions in the parameter space} for the well-motivated gauged \lmultau model. This is a simple anomaly-free extension of the SM which feature a $Z^\prime$ boson that mediates new interactions in the neutrino sector. However, our results can readily be applied also for other models with light mediators, such as the models for large Non-Standard neutrino Interaction (NSI) with matter fields \cite{Farzan:2019xor,Denton:2018dqq,Denton:2018xmq,Farzan:2017xzy,Farzan:2016wym,Farzan:2015hkd,Farzan:2015doa}.

This article is organised as follows. In \cref{sec:models}, we review the well-known \lmultau model, which incorporates a new vector coupling to the SM. In \cref{sect:SMproto}, we summarise the neutrino diffusion process in a proto-NS when only SM interactions are considered. In \cref{sec:impact}, we consider new physics contributions to neutrino-antineutrino scattering, including the production of the mediator on-shell, as well as the radial dependence of the density and temperature inside the proto-NS. We study the effects in two complementary regions of values for the gauge coupling. \cref{sec:SN-large} addresses the large coupling regime, for which we compute the regions of the parameter space where the neutrino-antineutrino scattering dominates over the SM neutrino-nucleon scattering contribution. We estimate the relevance of the neutrino energy redistribution and discuss that it is not discernible with the SN~1987A data. In \cref{sec:SN-small}, we study the small coupling regime, where the gauge mediator can decay in a distance longer than the neutrino-nucleon mean free path, leading to a shorter neutrino burst. Based on the observed duration of the SN~1987A neutrino signal, we \PRDnote{find} new \PRDnote{disfavoured} regions of the parameter space where the mediator decay length is smaller than the size of the proto-NS in the \lmultau model. Our conclusions are presented in \cref{sec:conclusions}.

\section{The \lmultau model}
\label{sec:models}

The recent measurement of the muon anomalous magnetic dipole moment, $(g-2)_\mu$, by the E989 experiment \cite{Muong-2:2021ojo} has confirmed a previously observed \cite{Bennett:2002jb,Bennett:2004pv,Bennett:2006fi} deviation with respect to the SM prediction. This can be interpreted as a hint of new physics in the leptonic sector. Among the different beyond the SM (BSM) realisations that could account for such an excess, the \lmultau gauge anomaly-free model is the simplest extension of the SM that would explain this observation introducing a light vector boson \cite{Baek:2001kca}. Moreover, such a model can offer an explanation for the nearly maximum mixing angle between the second and third generations of neutrinos \cite{Ma:2001md}, alleviate the tension between the late and the early time determinations of the Hubble constant \cite{Escudero:2019gzq} and accommodate dark matter with the correct relic abundance \cite{Biswas:2016yan,Altmannshofer:2016jzy,Gninenko:2018tlp, Kamada:2018zxi, Foldenauer:2018zrz}.

Due to the extremely small coupling to electrons or quarks, the \lmultau vector boson is difficult to test in collider or fixed target experiments. However, experiments looking for new physics in the neutrino sector together with SN and NS signals offer an opportunity to search for this new vector boson exploiting its tree-level couplings to the muon and tau neutrinos. Recent works \cite{Escudero:2019gzq, Kamada:2015era, Hardy:2016kme, Huang:2017egl, Croon:2020lrf} have provided constraints over the model by studying its cosmological and astrophysical implications. Bounds on the effective number of relativistic degrees of freedom, $N_{\rm eff}$, from Big Bang nucleosynthesis \cite{Kamada:2015era, Huang:2017egl} and stellar cooling \cite{Hardy:2016kme} strongly constrain new vector bosons with masses below 1~MeV. However, there is still room for heavier mediators which can both explain the $(g-2)_\mu$ measurement and be produced in the interior of the proto-NS core.

The Lagrangian of this model can be expressed as
\begin{equation}
  \mathcal{L}_{L_{\mu}-L_{\tau}}=\mathcal{L}_{\mathrm{SM}}-\frac{1}{4} Z^{\prime \alpha \beta} Z_{\alpha \beta}^{\prime}+\frac{m_{Z^{\prime}}^{2}}{2} Z_{\alpha}^{\prime} Z^{\prime \alpha}+Z_{\alpha}^{\prime} J_{\mu-\tau}^{\alpha} ,
\end{equation}
where $m_{Z^{\prime}}$ is the mass of the gauge boson, and $Z_{\alpha \beta}^{\prime} \equiv \partial_{\alpha} Z_{\beta}^{\prime}-\partial_{\beta} Z_{\alpha}^{\prime}$ is the field strength tensor. The $\mu-\tau$ current is
\begin{equation}
  J_{\mu-\tau}^{\alpha}=g_{\mu-\tau}\left(\bar{\mu} \gamma^{\alpha} \mu+\bar{\nu}_{\mu} \gamma^{\alpha} P_{L} \nu_{\mu}-\bar{\tau} \gamma^{\alpha} \tau-\bar{\nu}_{\tau} \gamma^{\alpha} P_{L} \nu_{\tau}\right),  
\end{equation}
where $P_{L}=\frac{1}{2}\left(1-\gamma_{5}\right)$ is the left chirality projector and $g_{\mu-\tau}$ is the gauge coupling.
The rest frame partial widths for $Z^{\prime}$ decays to charged leptons $\beta=\mu, \tau$ and neutrinos $\nu_{\beta}=\nu_{\mu, \tau}$ can be written as
\begin{align}
  \Gamma_{Z^{\prime} \rightarrow \beta^{+} \beta^{-}}&=\frac{g_{\mu-\tau}^{2} m_{Z^{\prime}}}{12 \pi}\left(1+\frac{2 m_{\beta}^{2}}{m_{Z^{\prime}}^{2}}\right) \sqrt{1-\frac{4 m_{\beta}^{2}}{m_{Z^{\prime}}^{2}}}\ , \\
  \quad \Gamma_{Z^{\prime} \rightarrow \bar{\nu}_{\beta} \nu_{\beta}}&=\frac{g_{\mu-\tau}^{2} m_{Z^{\prime}}}{24 \pi}\ . 
  \label{DW}
\end{align}
Note that for $m_{Z'} < 2 m_{\mu} \sim 211.3$ MeV, the only possible decay of the mediator is into $\mu$ and $\tau$ neutrinos. In our analysis, we will consider $1 \, \textrm{MeV} \leq m_{Z'} \leq 1$ GeV. The decay into muons will take place only for $Z'$ heavier than twice the muon mass. For $m_{Z'} \sim 1$ GeV, $\Gamma_{Z^{\prime} \rightarrow \mu^{+} \mu^{-}}/ \left(\Gamma_{Z^{\prime} \rightarrow \bar{\nu}_{\mu} \nu_{\mu}}+ \Gamma_{Z^{\prime} \rightarrow \bar{\nu}_{\tau} \nu_{\tau}}+\Gamma_{Z^{\prime} \rightarrow \mu^{+} \mu^{-}} \right) \sim 0.5$.

\section{Standard picture of neutrino diffusion in a proto-NS }
\label{sect:SMproto}

The diffusion of neutrinos from the core of proto-NS is a complicated process. Let us first focus on the diffusion of $\nu_\mu$, $\nu_\tau$, $\bar{\nu}_\mu$ and $\bar{\nu}_\tau$ (collectively denoted as $\nu_x$) within the SM. During the short $\nu_e$-burst and accretion periods ($t<0.7$~s), the $\nu_e$ and $\bar{\nu}_e$ fluxes are much larger than the $\nu_x$ flux but, in the proto-NS cooling period, the luminosities in the form of all three neutrino flavors and their antiparticles become equal up to a precision of 10\% \cite{Janka:2017vlw-book}. The total binding energy of the proto-NS (from the binding energy of the collapsing star) is of the order of $10^{53}$~erg. This energy will be depleted with a grey body radiation of neutrinos and antineutrinos of all three flavors from the neutrinosphere (at a radius of $\sim 20$~km). At the onset of the cooling phase ($\sim 1$~s after the bounce), the luminosity is of the order of $10^{52}$~erg~s$^{-1}$ for each neutrino and antineutrino species. However, as the outer shells cool down fast, the neutrinosphere recedes to smaller radii and the luminosity quickly drops. The neutrino emission is backed up with the diffusion of neutrinos from the inner shells. Due to multiple scattering, these neutrinos take a sizable time to reach the outer shells of the proto-NS, from where they are radiated out with a time scale of \cite{Janka:2017vlw-book} 
\begin{equation} 
  t_E\sim \frac{3 }{\pi^2 }\frac{\mathcal{E}_{th}^{tot}}{2\mathcal{E}_{th}^\nu} 
  R_{ns}^2\left\langle \frac{1}{\lambda_{\nu}}\right\rangle, 
  \label{eq:tE}
\end{equation}
where $\mathcal{E}_{th}^{tot}$ and $\mathcal{E}_{th}^\nu$ are respectively the total baryon and neutrino thermal energies, $\langle 1/\lambda_{\nu}\rangle$ is the average of the inverse of the neutrino mean free path. Remember that $R_{ns}^2\langle {1}/{\lambda_{\nu}}\rangle$ (where $R_{ns}$ is the radius of the neutrinosphere) gives the time scale of the diffusion of a single particle with a velocity of light and with random walk steps of $\lambda_{\nu}$. Within the SM, $\nu_x$ scattering is dominated by neutral current scattering on nucleons. The cross section is proportional to $\langle E_\nu^2\rangle$, which varies across the core radius.

The neutrino interaction rate with other particles in the interior of the proto-NS, as well as its mean free path and diffusion time, depend on the nuclear medium properties, such as the baryonic density, $n_B$, temperature, $T$, the effective nucleon masses, $m_{N}^{\star}$, and the neutrino and nucleon effective chemical potentials, $\mueff_{\nu_e}$, $\mueff_n$, $\mueff_p$. Note that the effective masses and chemical potentials of the particles in the nuclear medium are different from the naked values in vacuum by the presence of meson fields \cite{glendenning}\footnote{In order to obtain the electron effective chemical potential, the equilibrium equation that must be solved involves effective meson fields. In particular, $\mueff_n + \mueff_{\nu_e}= \mueff_p+\mueff_e +2 g_{\rho}\langle \rho\rangle$, where $\rho$ is an effective field responsible of the strong interaction in the model used in Ref.~\cite{Cerdeno:2021cdz}.}.

In \cref{tab:shells}, the temperatures, densities, and lepton fractions for the different shells of the star for a time of $1$~s after bounce, obtained from the simulation of Ref.~\cite{fischer2012}, are shown. The effective electron neutrino and nucleon chemical potentials and the effective nucleon masses were derived in Ref.~\cite{Cerdeno:2021cdz} invoking the TM1 model, widely used in current numerical simulations. This was done for a $18 \,M_\odot$ progenitor in a relativistic mean field approach, after solving the equations of motion and imposing the equilibrium conditions. Note that since the mass of the tau leptons is considerably larger than the temperature in the core, charged-current interactions for the $\tau$ neutrinos cannot take place. Thus, $\nu_\tau$ and $\bar{\nu}_\tau$ will be produced with equal amount \cite{Janka:2017vlw-book}. This implies $ \mu_{\nu_\tau}=\mu_{\bar{\nu}_\tau}=0$, unless weak magnetism effects \cite{horo2002} are considered. Equally, the net muon number in the proto-NS is negligible and, therefore, to a first approximation, we can also set $\mu_{\nu_\mu}\simeq \mu_{\bar{\nu}_\mu}\simeq 0$ \cite{Janka:2017vlw-book}. With a Fermi-Dirac energy distribution, the mean energy of $\nu_\mu$ and $\bar{\nu}_\mu$ is then $\left< E_\nu \right> \sim \pi T$, while for the electron neutrinos $\left< E_\nu \right> =(3/4) \mueff_\nu $. 

\begin{center} 
  \renewcommand{\arraystretch}{1.2}
  \begin{table}[t!]
  \begin{tabular}{l|c|c|c|c||c|c|c|c}
  \multicolumn{4}{l}{ \rule{0ex}{2.6ex}} \\ 
  \hline
  \hline
  & $R$ (km) &$T$ (MeV) & $n_B (\rm fm^{-3})$ & $Y_e$&$\mueff_n$ (MeV) & $\mueff_p$ (MeV) & $\mueff_{\nu_e}$ (MeV)& $m_{N}^{\star}$ (MeV) \rule{0ex}{2.6ex}
  \\ \hline
  $k=1$& 5.0&
  15 & 0.5 & 0.3
  & 496.6 & 405.4 & 114.6 & 249.6 
  \\ \hline
  $k=2$& 7.5&
  20 & 0.3 & 0.28
  & 530.0 & 458.3 & 102.7 & 384.9 
  \\ \hline
  $k=3$& 10.0&
  28& 0.15 & 0.25
  & 656.5 & 601.9 & 79.9 & 599.4 
  \\ \hline
  $k=4$& 15.0&
  33& 0.06 & 0.2
  & 779.8 & 723.0 & 29.0 & 786.0 
  \\ \hline
  $k=5$& 17.5&
  18& 0.03 & 0.1
  & 858.7 & 813.1 & 14.4 & 857.0 
  \\ \hline
  $k=6$& 20.0&
  7& 0.008 & 0.05
  & 917.2 & 893.9 & 12.5 & 915.9 
  \\ 
  \hline
  \hline
  \end{tabular}
  \caption{Values of neutron effective chemical potential, $\mueff_n$, proton effective chemical potential, $\mueff_p$, electron neutrino effective chemical potential, $\mueff_{\nu_e}$, and nucleon effective mass, $m_{N}^{\star}$, for the spherical shells (labeled by the index $k$ and defined by an outer radius $R_k$) that we consider at $1$ s after bounce, with a baryonic density, $n_B$, temperature, $T$ and electron fraction, $Y_e$. Temperatures, densities and electron fraction are taken from Ref.\,\cite{fischer2012}.
  }
\label{tab:shells}
\end{table}
\end{center}

\begin{figure}[!t]
  \centering
  \includegraphics[width=.45\textwidth]{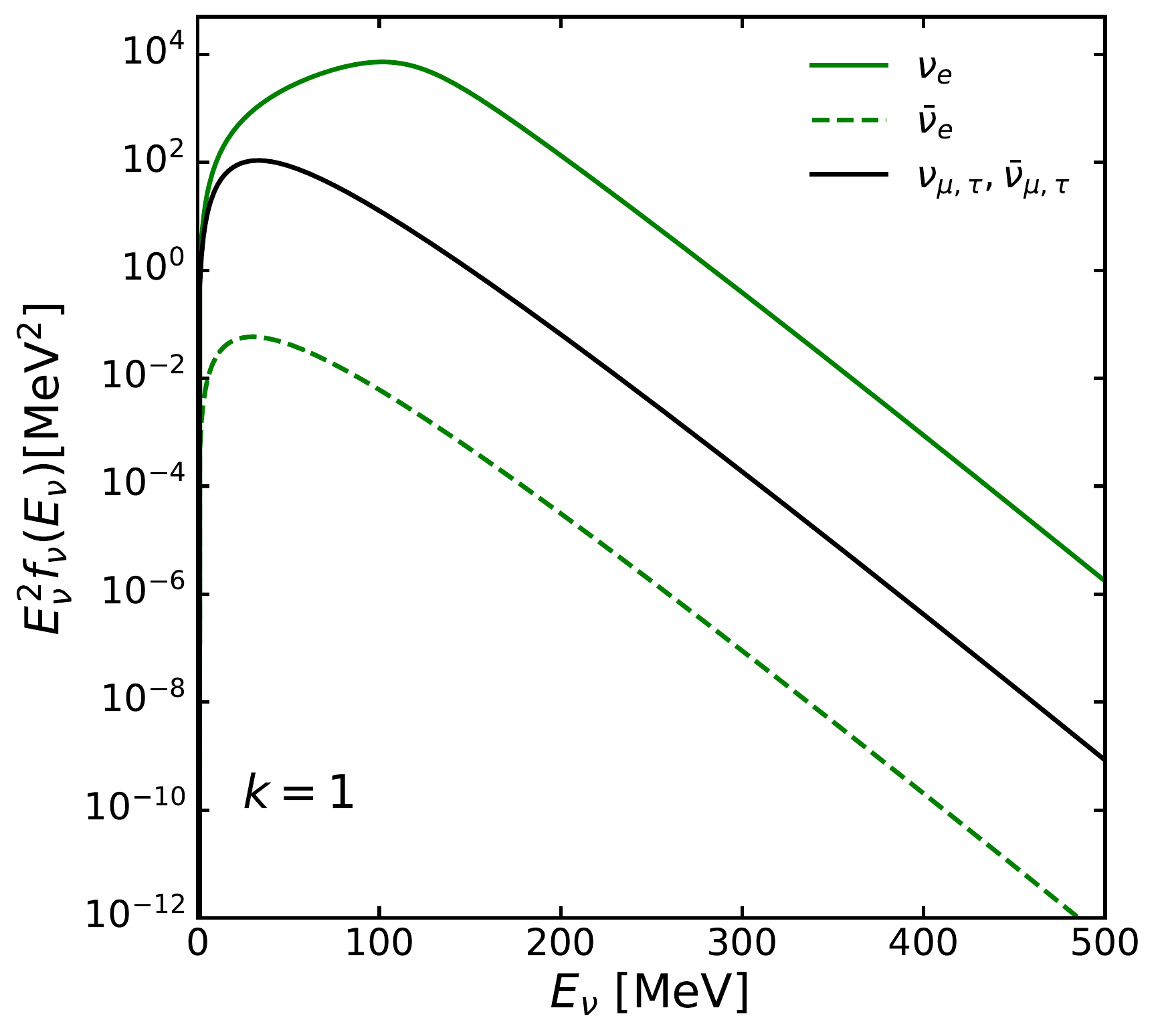}
  \includegraphics[width=.45\textwidth]{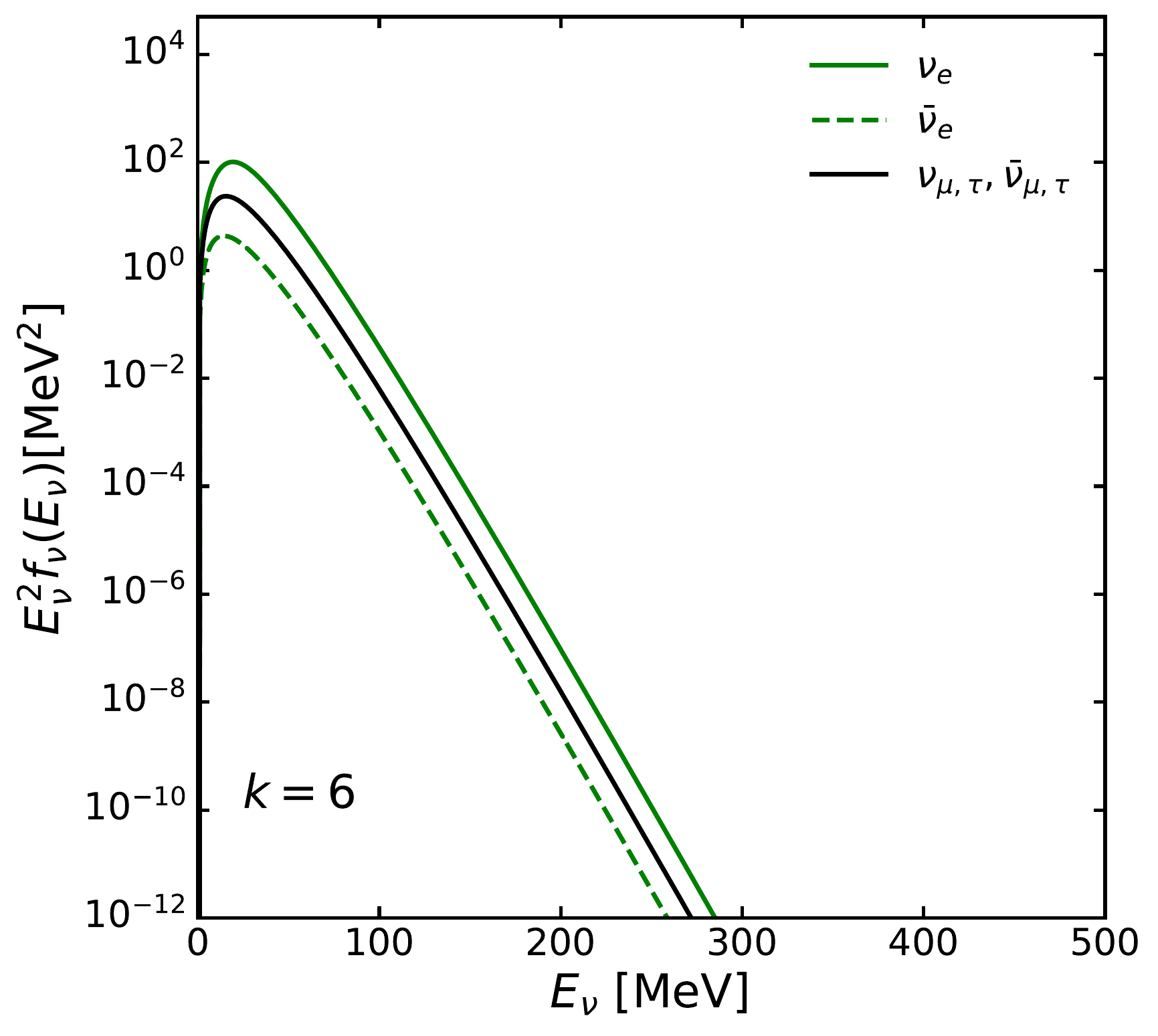}
  \caption{Energy distribution of electron (green) and muon/tau (black) neutrinos (solid) and antineutrinos (dashed) as a function of the neutrino energy for shells $k=1$ (left) and $k=6$ (right) of \cref{tab:shells}.
  }
  \label{fig:fnu}
\end{figure}

In \cref{fig:fnu}, we have represented the energy distribution, $E_{\nu_\beta}^2 f(E_{\nu_\beta}, \mueff_{\nu_\beta}, T) $, for muon and tau neutrinos (black) and for electron neutrinos (green solid) and antineutrinos (green dashed), as a function of the neutrino energy for the first and sixth shells of the star (see \cref{tab:shells}), on the left and right plots respectively. 
Since the chemical potential for muon/tau neutrinos and antineutrinos is negligible, their distribution functions coincide, $f_{\nu_{\mu,\tau}}=f_{\bar\nu_{\mu,\tau}}$. On the contrary, for electron antineutrinos $\mueff_{\bar{\nu}_e}=-\mueff_{\nu_e}$ and, therefore, in the inner shells, where the electron neutrino chemical potential reaches the highest values, their distribution functions substantially differ, $f_{\nu_e}\gg f_{\bar\nu_e}$. This, in turn, leads to a negligible density of electron antineutrinos in the inner shells, which suppresses the $\nu_e \bar{\nu}_e$ interactions. In the outer shells, the electron neutrino and antineutrino chemical potentials significantly decrease in absolute value and therefore $f_{\nu_e}\sim f_{\bar\nu_e} \sim f_{\nu_{\mu,\tau}}$. Due to the high matter density and frequent collisions with the SN medium inside the proto-NS, flavour conversions are expected to be suppressed. Neutrino oscillation becomes possible only after neutrinos exit the core.

The measurement of SN 1987A neutrino burst duration is subject to uncertainties, especially due to the unknown relative time offsets (of up to $\pm1$ minute) of the three detectors, Kamiokande II, IMB, and Baksan. However, from the Kamiokande II observations alone, we can make sure that the duration cannot be shorter than 10~s. For example, a time duration of 5~s is already ruled out. The constraints that we are going to set in this paper are based on the shortening of the duration which is robust against the measurement uncertainties. We should however notice that constraining a model that prolongs the duration by a factor of $\sim 2$ can be challenged by the uncertainty.

The duration of the neutrino signal in the SM can be computed assuming that neutrino-nucleon interactions are the leading contribution to the neutrino mean free path, $\lambda_{\nu_\beta}$. The neutrino diffusion time can be computed as
\begin{equation}
   t^{\nu_\beta}_{\rm diff} = \sum_{k=1}^n \left( R_k^2-R_{k-1}^2\right) \left<1/\lambda_{\nu_{\beta}}\right>_k, 
  \label{eq:deltat}
\end{equation}
where $k=1,...,6$ are the different shells of \cref{tab:shells} at 1~s after the bounce, $R_{k}$ are the radius of each shell, and $\left<1/\lambda_{\nu_{\beta}}\right>_k$ is the average of the inverse of the mean free path in each shell, 
\begin{equation}
  \langle 1/\lambda_{\nu_{\beta}}\rangle = \frac{\int d E_{\nu_{\beta}} f(E_{\nu_{\beta}}, \mueff_{\nu_\beta}, T ) E_{\nu_{\beta}}^{2} \lambda_{\nu_{\beta}}^{-1}(E_{\nu_{\beta}})}{\int d E_{\nu_{\beta}} f(E_{\nu_{\beta}}, \mueff_{\nu_\beta}, T ) E_{\nu_{\beta}}^{2}},
\end{equation}
where $E_{\nu_\beta}$, $E_{N}$, $\vec{p}_{\nu_\beta}$, $\vec{p}_{N}$ are the energies and momenta of the incoming neutrino and nucleon, $E'_{\nu_\beta}$, $E'_{N}$, $\vec{p'}_{\nu_\beta}$, $\vec{p'}_{N}$ are those of the outgoing states, and $p_{\nu_\beta}$, $p_{N}$, $p'_{\nu_\beta}$, $p'_{N}$ are the corresponding four-momenta. In this expression, $\beta=e, \mu, \tau$ indicates the neutrino flavour, $N=n, p$ refers to the nucleon states, {\it i.e.}, neutrons and protons, and
\begin{equation}
  \lambda_{\nu_{\beta}}^{-1}(E_{\nu_{\beta}})=\sum_N \int 2 \frac{d^3 \vec{p_N}}{(2 \pi)^3} f(E_N, \mueff_N, T) |\vec{v_{\nu_\beta}}-\vec{v_{N}}| \sigma_{\nu_\beta, N}.
  \label{eq:mfp}
\end{equation}
In this expression, the product of the neutrino-nucleon relative velocity and the neutrino-nucleon scattering cross section is given by
\begin{equation}
	|\vec{v_{\nu_\beta}}-\vec{v_{N}}| \sigma_{\nu_\beta, N}=\int d\Phi_{\nu_\beta, N}\,  \frac{|\mathcal{\overline{M}}|_{\nu_{\beta}, N}^2}{4 E_{\nu_\beta} E_N}\mathcal{F}(E'_\nu, E'_{N}),
\end{equation}
where
\begin{equation}
d\Phi_{\nu_\beta, N}=  \frac{d^3 \vec{p'}_{\nu_\beta}}{(2 \pi)^3 2 E'_{\nu_{\beta}}} \frac{d^3 \vec{p'}_{N}}{(2 \pi)^3 2 E'_{N}} (2 \pi)^4 \delta^{(4)}(p_{\nu_\beta}+p_N-p'_{\nu_\beta}-p'_N)  
\end{equation}
is the 2-body neutrino-nucleon phase space element.
$f(E_{N}, \mu^*_{N}, T))$ is the Fermi-Dirac distribution function for the incoming nucleon and $\mathcal{F}(E'_\nu, E'_{N})=(1- f(E'_\nu, \mu^*_\nu, T))(1- f(E'_{N}, \mu^*_{N}, T))$ accounts for the Pauli blocking of the outgoing states.

If we consider only the SM interactions, we obtain $t^{\nu_{\mu,\tau}}_{\rm diff} \sim 1.3 \, \rm s$ for muon and tau neutrinos and $t^{\nu_{e}}_{\rm diff} \sim 3 \, \rm s$ for electron neutrinos. This leads to $t_E\sim 10$~s in \cref{eq:tE}, due to the fact that the stored thermal energy in matter continues to be emitted in neutrinos even after the first neutrinos escape, which is compatible with the observed duration of the neutrino signal from SN~1987A.

\section{Impact of on-shell production of new light mediators on neutrino diffusion}
\label{sec:impact}

In this section, we analyse two complementary regimes: when the new gauge coupling is large (so that the neutrino-antineutrino interaction strength is comparable or exceed the SM neutrino-nucleon one) and when it is small (so that the new mediator can travel a long distance inside the proto-NS star before decaying back to neutrinos).

\begin{figure}[!t]
  \centering
  \includegraphics[width=.31\textwidth]{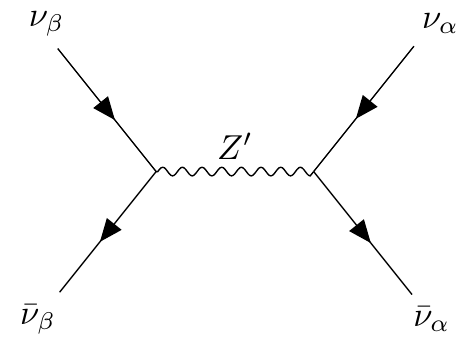}\hspace*{1cm}
  \includegraphics[width=.31\textwidth]{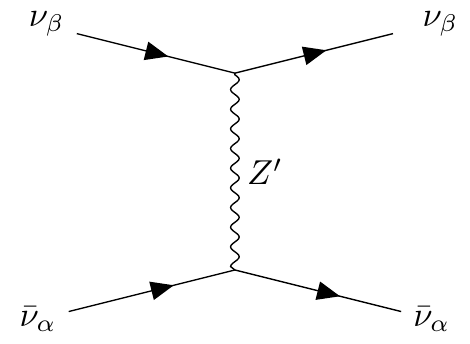}
  \caption{
  New physics contribution to neutrino-antineutrino scattering through a low-mass vector boson, $Z^\prime$. The diagram on the left represents the $s$-channel that is the leading contribution for the relevant range of parameters. For the \lmultau model $\alpha, \, \beta=\mu,\, \tau$. 
  } 
  \label{fig:diagram_ann}
\end{figure}

In \cref{fig:diagram_ann}, we show the neutrino-antineutrino scattering diagrams via the new vector mediator, $Z'$. The cross section of this process can be greatly enhanced through $s$-channel Breit-Wigner resonance when the mediator is produced on-shell\footnote{When $Z'$ reaches thermal equilibrium with the neutrino gas and the plasma, it can receive a share of the entropy content of the core. We may therefore wonder whether this significantly reduces the temperature relative to the SM prediction. Since neutrinos below the so-called energy neutrinosphere are in thermal equilibrium with the plasma, which act as a huge thermal energy source ({\it i.e.}, $\mathcal{E}_{th}^{tot}/\mathcal{E}^{\nu}_{th} \gg 1$), the addition of the three bosonic degrees of freedom associated with the three polarizations of the $Z'$ boson can only change the temperature by a negligible amount suppressed by $\mathcal{E}^{\nu}_{th}/\mathcal{E}_{th}^{tot} \sim 0.2$. We therefore assume that around $1$~s after the bounce, the temperature profile of the core is similar to what is predicted within the SM even in the limit of large coupling. }. This is very sensitive to the energy distribution of neutrinos in the SN core of Fig.~\ref{fig:fnu}.

The $t$-channel contribution would be leading at very large values of the coupling (since it scales with the fourth power of the coupling) and small mediator masses (due to collinear enhancement), {\it i.e.}, $m_{Z'}\sim 1$ MeV for $g_{\mu - {\tau}} \sim 0.1$, however, these regions of the parameter space are generally ruled out by existing experimental and observational constraints. We have explicitly checked that, when the mediator goes on-shell, the main contribution to the neutrino-antineutrino cross section is given by the $s$-channel diagram. Likewise, other processes such as neutrino-neutrino scattering and neutrino-lepton scattering, both of which would proceed through $t-$channel, are sub-leading.

The $s$-channel Breit-Wigner resonance in neutrino-antineutrino scattering is not relevant in the SM, since the typical neutrino energies inside the proto-NS are not high enough to produce the $Z$ boson on-shell. However, in models with new gauge bosons in the MeV range, the Breit-Wigner resonance can significantly increase the neutrino-antineutrino scattering cross section. The model that we consider is lepton flavor conserving so $\nu_\beta$ will fuse only with $\bar{\nu}_\beta$ to produce an on-shell $Z'$.

\subsection{Large coupling regime}
\label{sec:SN-large}

Let us first discuss how neutrino-antineutrino interactions can affect the neutrino burst duration in the relatively large coupling regime, where the rate of neutrino-antineutrino interactions is comparable or larger than the neutrino-nucleon SM interaction rate.

The impact of scatterings of neutrinos off nucleons on the average diffusion time is quite different from the effects of neutrino scattering off the background antineutrinos. An outflow of neutrinos will be hindered by the scattering off nucleons and electrons inside the core. In the following, we show the difference using the formalism in Ref.~\cite{Pascal:2022qeg}. The time evolution due to the neutrino and antineutrino outflow is given by the radial derivation of the neutrino-antineutrino energy flux \cite{Pascal:2022qeg}: $4\pi \int(H_\nu+H_{\bar{\nu}})E^3 dE$, where $H_{\nu (\bar{\nu})} =\int_{-1}^{+1} f_{\nu(\bar{\nu})} \cos\theta_r d\cos\theta_r$ in which $f_{\nu(\bar{\nu})}$ is the neutrino (antineutrino) distribution and $\theta_r$ is the angle that momentum of the neutrino (antineutrino) makes with the radial direction. Because of the net outflow of neutrinos and antineutrinos, $H_\nu$ and $H_{\bar{\nu}}$ are nonzero. That is, since neutrinos and antineutrinos move more frequently outward than inwards, even isotropic scattering on nucleons or electrons can reduce $H_\nu+H_{\bar{\nu}}$. However, when neutrinos and antineutrinos scatter off each other as $\nu+\bar{\nu} \to \nu+\bar{\nu}$, $H_\nu+H_{\bar{\nu}}$ remains unchanged because this sum is nothing but the total momentum density in the radial direction, which is conserved. In case of scattering with the electrons and nucleons, this sum can change because the neutrino and antineutrinos exchange momentum with the plasma.

However, in this limit, the neutrino-antineutrino coalescence can in principle redistribute neutrino energies with a rate larger than that of scattering off the background matter, which brings the distribution back to the thermal Fermi-Dirac distribution. Let us consider a neutrino with energy $E_1\sim T$ scattering off an antineutrino whose momentum makes an angle of $\theta$ with the direction of the initial neutrino. In order for such a pair to produce an on-shell $Z'$, the antineutrino energy should be 
\begin{equation}
  E_2 = \frac{m_{Z'}^2}{2E_1 (1-\cos\theta)}.
  \label{eq:e2}
\end{equation}

Simple kinematics\footnote{If $Z'$ scatters before decay, its energy will change and this argument does not therefore apply. Thermalization of $Z'$ however requires $\sigma(Z'\nu \to Z' \nu) n_\nu/\Gamma_{Z'}>1$ which can be achieved only for couplings larger than 1. } show that the final neutrino and antineutrino from the $Z'$ decay will have flat energy distribution in the range $[(E_1+E_2)(1-v_{Z'})/2,(E_1+E_2)(1+v_{Z'})/2]$ in which $v_{Z'}=(1-m_{Z'}^2/(E_1+E_2)^2)^{1/2}$. For $E_2\sim E_1 \sim T$, the energies of the final particles will be of the same order as those of the initial neutrinos and, therefore, their diffusion time cannot be significantly altered. In the limit $E_2\gg E_1$, given that the scattering cross section of neutrinos off nuclei is proportional to the square of the neutrino energy, the diffusion time can change. If the neutrino-antineutrino interaction rate is small, the impact will of course be tiny. Let us take $E_{lim}$ to be an arbitrary value with the condition $E_{lim}\gg T$. On the other hand, in case the rate of scattering off antineutrinos (or neutrinos) with energy $E_2$ such that $E_2> E_{lim}\gg T$ is larger than the rate of scattering off stellar matter (which thermalises back the energy distribution), the higher energy tail of the neutrino distribution (with energies exceeding $E_{lim}$) will be used up, without having time to be replaced. Thus, independently of the value of the coupling, the fraction of neutrinos or antineutrinos that come out of the thermal Fermi-Dirac distribution will be very tiny and given by the ratio of the number density of antineutrinos with energy larger than $E_{lim}$ ($n_{lim}$) to the total number density of neutrinos. This can be understood with the following argument. The dynamics of $n_{lim}$ can be described as $\dot{n}_{lim}=-\Gamma_{SM}(n_{lim}-n_{lim}^{eq})-\Gamma_{NEW} n_{lim}$ where $\Gamma_{SM}$ is determined by the rate of scattering off electrons and nucleons and $n_{lim}^{eq}$ is the value of $n_{lim}$ computed with the Fermi-Dirac distribution $n_{lim}^{eq}=\int_{E_{lim}}^\infty f(E_\nu,-\mueff_\nu,T) E_\nu^2 dE_\nu$. The asymptotic (stable) value of $n_{lim}$ will be given by $n_{lim}^{asym}=n_{lim}^{eq} \Gamma_{SM}/\Gamma_{NEW}$. Let us denote the number density of neutrinos scattered up to higher energies out of the Fermi-Dirac distribution by $n^{out}$. The evolution of $n^{out}$ can be written as $\dot{n}^{out}=-\Gamma'_{SM} n^{out}+\Gamma_{NEW} n_{lim}^{asym}$. Thus, the asymptotic solution which shows the density of neutrinos kicked out of equilibrium is $n^{out}=(\Gamma_{SM}/\Gamma_{SM}') n_{lim}^{eq}$. For $E_{lim}\gg T$, this quantity is suppressed by a factor
\begin{equation}
  \frac{\int_{E_{lim}}^\infty f(E_\nu,\mueff_\nu,T)E_\nu^2 dE_\nu}{\int_{0}^\infty f(E_\nu,-\mueff_\nu,T)E_\nu^2 dE_\nu}\ll 1 .
\end{equation}

A less extreme scenario is the one where $E_{ 1} = \pi T - \Delta$ and $E_{ 2}$ as in \cref{eq:e2}, where $0<\Delta< \pi T$ is an arbitrary energy. For example, let us consider muon and tau neutrinos with energies of $E_{1}=15$~MeV, interacting with antineutrinos with energies $E_{2}=m_{Z'}^2/(30 \, \textrm{MeV} (1-\textrm{cos}\,\theta))$. If the mediator mass is $m_{Z'}=50$ MeV, the antineutrino energies must be $E_{2} \gtrsim41.7$~MeV in order to produce the mediator on-shell. After the interaction, $\nu_1$ will gain energy and therefore its interactions with nucleons will be stronger, since $\sigma_{\nu, N} \propto E_{\nu}^2$. As a result, $\nu_1$ will be more bounded to stellar matter and stay for longer in the star.

The lower energy tail of the spectrum can therefore be scattered up to higher energies via $Z'$ on-shell production. We should however remember that the fraction of muon or tau neutrinos with $E_\nu<T$ ($E_\nu<T/3$) is only 8\% (0.48\%) of the whole number density. Thus, even if all of neutrinos in the low energy tail scatter up, the impact on the diffusion time is not observable in the SN~1987A data. A more detailed analysis is needed to determine whether this can be a noticeable feature in the event of future SN detection with a greatly improved statistics.

In the following subsection, we study the regions of the parameter space where this effect could take place for the well-motivated \lmultau model.

\subsubsection{Implications for the parameter space of the \lmultau model}
\label{sub:models}

To find the regions where neutrino-antineutrino interactions are more frequent than neutrino-nucleon interactions, we need to calculate the neutrino-antineutrino scattering rate via the new $Z'$ mediator. The average of the neutrino-antineutrino scattering rate can be calculated by integrating the scattering cross section as follows,
\begin{equation}
	\langle \mathcal{R}_{\nu_{\beta} \bar\nu_{\beta} \to Z' \to \nu_{\alpha} \bar \nu_{\alpha}}\rangle= \frac{\int d E_{\nu_{\beta}} f(E_{\nu_{\beta}}, \mueff_{\nu_\beta}, T ) E_{\nu_{\beta}}^{2} \mathcal{R}_{\nu_{\beta} \bar\nu_{\beta} \to Z' \to \nu_{\alpha} \bar \nu_{\alpha}}(E_{\nu_{\beta}})}{\int d E_{\nu_{\beta}} f(E_{\nu_{\beta}}, \mueff_{\nu_\beta}, T ) E_{\nu_{\beta}}^{2}},
	\label{eq:avmfp}
\end{equation}
where $\beta= \mu, \tau$ indicates the neutrino flavour and
\begin{equation}
	\mathcal{R}_{\nu_{\beta} \bar\nu_{\beta} \to Z' \to \nu_{\alpha} \bar \nu_{\alpha}}(E_{\nu_{\beta}})=\int \frac{d^3 \vec{p}_{\bar{\nu}_\beta}}{(2 \pi)^3} f(E_{\bar{\nu}_\beta}, \mueff_{\bar{\nu}_\beta}, T) |\vec{v}_{\nu_\beta}-\vec{v}_{\bar{\nu}_\beta}| \sigma_{\nu_\beta, \bar{\nu}_\beta}
\end{equation}
is the neutrino-antineutrino scattering rate via $Z'$, $|\vec{v}_{\nu_\beta}-\vec{v}_{\bar{\nu}_\beta}|$ is the relative velocity between neutrinos and antineutrinos and $\sigma_{\nu_\beta, \bar{\nu}_\beta}$ is the neutrino-antineutrino scattering cross section. Note that we are considering massless neutrinos and therefore $|\vec{p}_{\nu_\beta}|=E_{\nu_\beta}$, $|\vec{p}_{\bar \nu_\beta}|=E_{\bar \nu_\beta}$.

For the $\nu_{\beta} \bar \nu_{\beta} \to Z' \to \nu_{\alpha} \bar \nu_{\alpha}$ process, the factor $|\vec{v}_{\nu_\beta}-\vec{v}_{\bar{\nu}_\beta}| \sigma_{\nu_\beta, {\bar{\nu}_\beta}}$ can be written as
\begin{equation}
	|\vec{v}_{\nu_\beta}-\vec{v}_{{\bar{\nu}_\beta}}| \sigma_{\nu_\beta, {\bar{\nu}_\beta}}=\int d\Phi_{\nu_\beta, \bar \nu_\beta}\, \frac{|\mathcal{\overline{M}}|_{\nu_{\beta}, {\bar{\nu}_\beta}}^2}{4 E_{\nu_\beta} E_{\bar{\nu}_\beta}}\mathcal{F}(E'_{\nu_\alpha}, E'_{\bar{\nu}_\alpha}),  
\end{equation}
where
\begin{equation}
 d\Phi_{\nu_\beta, \bar \nu_\beta}=  \frac{d^3 \vec{p'}_{\nu_\alpha}}{(2 \pi)^3 2 E'_{\nu_{\alpha}}} \frac{d^3 \vec{p'}_{\bar{\nu}_\alpha}}{(2 \pi)^3 2 E'_{\bar{\nu}_\alpha}} (2 \pi)^4 \delta^{(4)}(p_{\nu_\beta}+p_{\bar{\nu}_\beta}-p'_{\nu_\alpha}-p'_{\bar{\nu}_\alpha}) 
\end{equation}
is the 2-body phase space element,
with $E_{\nu_\beta}$, $E_{\bar{\nu}_\beta}$, $\vec{p}_{\nu_\beta}$, $\vec{p}_{\bar{\nu}_\beta}$ the energies and momenta of the incoming particles, $E'_{\nu_\alpha}$, $E'_{\bar{\nu}_\alpha}$, $\vec{p'}_{\nu_\alpha}$, $\vec{p'}_{\bar{\nu}_\alpha}$ those of the outgoing states, and $p_{\nu_\beta}$, $p_{\bar{\nu}_\beta}$, $p'_{\nu_\alpha}$, $p'_{\bar{\nu}_\alpha}$ the corresponding four-momenta. The factor $\mathcal{F}(E'_{\nu_\alpha}, E'_{\bar{\nu}_\alpha})=(1- f(E'_{\nu_\alpha}, \mu^*_{\nu_\alpha}, T))(1- f(E'_{\bar{\nu}_\alpha}, \mu^*_{\bar{\nu}_\alpha}, T))$ accounts for Pauli blocking in the outgoing states.

Since, in the parameter space under analysis, the on-shell production of the $Z'$ boson is the leading new physics process and ${\Gamma_{Z' \to \bar{\nu}_\beta \nu_\beta} }/{m_{Z'}}\ll 1$ is fulfilled, we can use the narrow width approximation (NWA). In this way, the neutrino-antineutrino interaction rate for these new physics interactions, $\mathcal{R}_{\nu_{\beta} \bar\nu_{\beta} \to Z' \to \nu_{\alpha} \bar \nu_{\alpha}}(E_{\nu_{\beta}})$, can be computed as 
\begin{equation}
  \mathcal{R}_{\nu_{\beta} \bar\nu_{\beta} \to Z' \to \nu_{\alpha} \bar \nu_{\alpha}}(E_{\nu_{\beta}})=\frac{1}{32 \pi} \int_{E^{min}_{\bar{\nu}_{\beta}}}^{\infty} dE_{\bar{\nu}_{\beta}} \frac{f(E_{\bar{\nu}_{\beta}}, \mueff_{\bar{\nu}_\beta}, T) E_{\bar{\nu}_{\beta}}}{E_{\bar{\nu}_{\beta}}+E_{\nu_{\beta}}} \left(\frac{m_{Z'}}{E_{\bar{\nu}_{\beta}} E_{\nu_{\beta}}} \right)^2  |\mathcal{\overline{M}}|_{ \nu_{\beta} \bar{\nu}_{\beta} \to Z'}^2 \frac{\Gamma_{Z' \to \nu_{\alpha} \bar{\nu}_{\alpha}}}{\Gamma_{Z'}^{tot}},
  \label{eq:lambdaNP}
\end{equation}
with $E^{min}_{\bar{\nu}_{\beta}}={m_{Z'}^2}/{(4 E_{\nu_{\beta}})}$ and $\Gamma_{Z'}^{tot}=\sum_{\alpha} \Gamma_{Z^{\prime} \rightarrow \bar{\nu}_{\alpha} \nu_{\alpha}}+\Gamma_{Z^{\prime} \rightarrow \beta^{+} \beta^{-}}$, where $\beta= \mu, \tau$. Note that the decay into taus will not be open for the values of the mediator masses considered in this work. The squared amplitude can be written as $|\mathcal{\overline{M}}|_{ \nu_{\beta} \bar{\nu}_{\beta} \to Z' }^2= g_{\mu-\tau}^2 m_{Z'}^2/2$. Here we are neglecting Pauli blocking for neutrinos and antineutrinos since for muon and tau neutrinos, which are the ones interacting with $Z'$, the chemical potential vanishes and therefore $(1- f(E'_{\bar \nu}, \mu^*_{\bar \nu}, T)) \sim 1$.

Within the SM, a suppressed population of muons is expected to exist inside a proto-NS. Ref.~\cite{Croon:2020lrf} has studied the $Z'$ production by this background muon population through semi-Compton and Bremsstrahlung processes, showing that the effect can be significant only for the $Z'$ masses below $\sim 5$~MeV, which are already ruled out by the cosmological bounds. We therefore neglect this effect. In principle, one can think that the annihilation of neutrino-antineutrino into muons via $s$-channel $Z'$ diagram may have an impact on the physics of the proto-NS. However, these processes generate the same amount of muons and antimuons, not changing their chemical potential, and therefore we do not expect the equation of state of the nuclear matter to change. Besides, even though muon neutrinos (antineutrinos) can scatter on antimuons (muons), as the muon density and its scattering cross section with muon neutrinos and antineutrinos are much smaller than that of nucleons, the mean free path of scattering on muons (antimuons) is negligible compared to the scattering on nucleons. The only effect we could expect comes from the decay of muons into muon neutrinos close to the neutrinosphere, which can change equality between the muon neutrino and tau neutrino fluxes that come out of the neutrinosphere, having consequences for collective neutrino oscillation \cite{Dasgupta:2007ws, Friedland:2010sc, Capozzi:2018ubv, Chakraborty_2020, Capozzi:2020kge}.

For completeness, we have also explicitly checked that processes at one loop, such as neutrino-antineutrino annihilation into $\mu^+\mu^-$ and the subsequent $\mu^+\mu^-\to\nu\bar\nu$ are sub-leading for the calculation of the neutrino-antineutrino scattering rate.

In this subsection, we show the parameter space from which the rate of neutrino-antineutrino interaction can exceed that of the neutrino scattering off nucleons for the \lmultau model. As we mentioned above, these regions cannot be ruled out using SN~1987A data, but they are indicative of the regions that might be explored when a better measurement of the SN neutrino flux (perhaps with better statistics from future detectors) is available. We shall compare them with the range ruled out with different observables and considerations.

Since in this scenario the new neutrino couplings to electrons and nucleons are extremely suppressed, the mean free path of electron neutrinos is not modified, and therefore charged-current $\beta$-processes do not need to be taken into account.

\begin{figure}[!t]
  \centering
  \includegraphics[width=.6\textwidth]{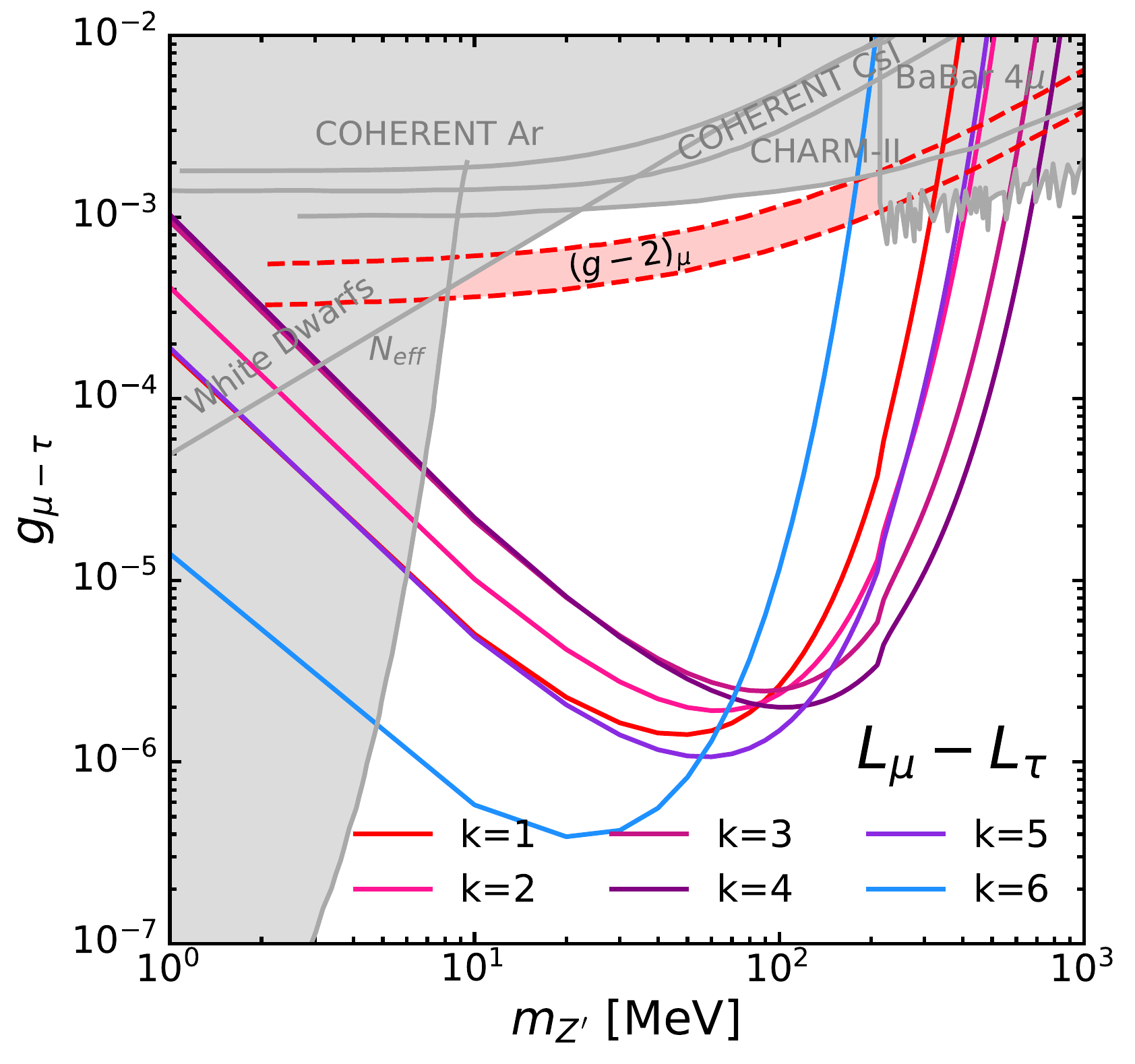}
  \caption{Values of $(g_{\mu-\tau},\,m_{Z^\prime})$ for which the average of the neutrino-antineutrino interaction rate of \cref{eq:avmfp} equals the SM value in each of the proto-NS shells of \cref{tab:shells}.
  The band in red shows the region in the parameter space consistent with the $(g-2)_\mu$ measurement. grey areas show different complementary bounds from: BaBar \cite{BaBar:2016sci}, neutrino tridents (Charm-II) \cite{Altmannshofer:2014pba, Chang:2022aas}, COHERENT CsI \cite{Abdullah:2018ykz} and LAr \cite{Amaral:2020tga}, white dwarf cooling \cite{Bauer:2018onh}, and the effect on $N_{eff}$ in the Early Universe \cite{Escudero:2019gzq}. 
  }
  \label{fig:bounds_lmultau}
\end{figure}

We have represented in \cref{fig:bounds_lmultau} the values of $g_{\mu-\tau}$ as a function of $m_{Z'}$ for which the new physics contribution to the neutrino scattering becomes as important as the SM one for each of the six shells described in \cref{tab:shells}. We should emphasize that, contrary to what might be inferred from Ref.~\cite{Kamada:2015era}, the region compatible with $(g-2)_\mu$ is not yet ruled out.

The inner shells of the proto-NS ($k=1-5$) are characterised by high temperatures and nucleon densities. These regions give the largest contribution to the neutrino scattering cross section both in the SM and when new physics is added. Due to the larger available neutrino energies, in these inner regions the condition for the light mediator production holds for larger masses (the tail of the neutrino distribution function - albeit very small - is still considerably larger than in the outer regions as we showed in \cref{fig:fnu}). From these lines, we see that a coupling of the order of $g_{\mu-\tau}\sim 10^{-4}-10^{-5}$ is sufficient for the new physics diagrams to be comparable to the SM contribution. A small kink at $ 2 m_\mu=210$~MeV can be observed in all the lines. This is due to the opening of the $Z^\prime\to\mu^+ {\mu^-}$ decay channel, which slightly increases the total $Z^\prime$ decay width in \cref{eq:lambdaNP}, reducing the cross section.

In the outer shell ($k=6$) the nucleon density is considerably smaller (the neutrinosphere is located at $\sim 20$~km). This leads to a large suppression of the SM processes (neutrino-neutron scattering) and makes it easier for the new physics to dominate (substantially reducing the values of $g_{\mu-\tau}$ at which this occurs). Also, the temperature is much smaller in these regions, which means that the Breit-Wigner resonant condition can only be fulfilled for small masses of $m_{Z^\prime}$. For this reason, the lines at which the new physics equals the SM contributions move to lower values of $g_{\mu-\tau}$ and $m_{Z^\prime}$. However, the time scale of neutrino diffusion from these two last layers up to the neutrinosphere is much smaller than $1$~s so they are not relevant for the diffusion time.

For comparison, we show in grey the existing bounds on the parameter space from various observations.

\subsection{Small coupling regime} 
\label{sec:SN-small}

In this section, we first compute the rate of $\nu \bar{\nu} \to Z'$ and the average decay length of $Z'$, considering the energy distribution of the produced $Z'$ and the relevant boost factor. We then focus on the parameter space range where $\ell_{Z'}$ is larger than the standard neutrino mean free path ($\sim 3~{\rm m}$) but smaller than the proto-NS radius (given by the size of the neutrinosphere, $R_{ns}\sim 20 $~km). This range has been overlooked in the literature. We also explore the range $\ell_{Z'} >R_{ns}$, confirming previous results. We shortly point out the potential effects on flavor and energy spectrum of neutrinos, arguing that they are distinct for the two regimes $\ell_{Z'}<R_{ns}$ and $\ell_{Z'}>R_{ns}$. Finally, we \PRDnote{identify new regions of the \lmultau model parameter space that are in tension with the SN 1987A neutrino burst duration measurement.}

If, after neutrino-antineutrino coalescence takes place, $Z'$ decays back into $\nu\bar\nu$, this would lead to neutrinos effectively escaping earlier, shortening the length of the SN neutrino burst. Similarly to \cref{eq:lambdaNP}, the rate of the scattering of a single $\nu$ off any $\bar{\nu}$ in the medium producing the $Z'$ on-shell can be computed via the following relation,
\begin{equation}
  \mathcal{R}_{\nu_{\beta} \bar\nu_{\beta} \to Z'}(E_{\nu_{\beta}})=\frac{1}{32 \pi} \int_{E^{min}_{\bar{\nu}_{\beta}}}^{\infty} dE_{\bar{\nu}_{\beta}} \frac{f(E_{\bar{\nu}_{\beta}}, \mueff_{\bar{\nu}_\beta}, T) E_{\bar{\nu}_{\beta}}}{E_{\bar{\nu}_{\beta}}+E_{\nu_{\beta}}} \left(\frac{m_{Z'}}{E_{\bar{\nu}_{\beta}} E_{\nu_{\beta}}} \right)^2  |\mathcal{\overline{M}}|_{ \nu_{\beta} \bar{\nu}_{\beta} \to Z' }^2 .
  \label{eq:ratenuanu}
\end{equation}
The time scale of the interaction can be defined as $\tau_{\nu_\beta\bar\nu_\beta\to Z'} ={c}/\langle\mathcal{R}_{\nu_{\beta} \bar\nu_{\beta} \to Z'}\rangle$, with
\begin{equation}
	\langle \mathcal{R}_{\nu_{\beta} \bar\nu_{\beta} \to Z' } \rangle= \frac{\int d E_{\nu_{\beta}} f(E_{\nu_{\beta}}, \mueff_{\nu_\beta}, T ) E_{\nu_{\beta}}^{2} \mathcal{R}_{\nu_{\beta} \bar\nu_{\beta} \to Z'}(E_{\nu_{\beta}})}{\int d E_{\nu_{\beta}} f(E_{\nu_{\beta}}, \mueff_{\nu_\beta}, T ) E_{\nu_{\beta}}^{2}}.
	\label{eq:avintrate}
\end{equation}

The decay length of $Z'$, $\ell_{Z'}$, is given by lifetime multiplied by the relativistic boost factor, $\gamma_{Z'}=E_{Z'}/m_{Z'}$, and the $Z'$ velocity, $v_{Z'}$, averaged over the energy of $Z'$, which is set by the energies of the neutrino and antineutrino producing the mediator on-shell,
\begin{equation}
  \ell_{Z'}=\frac{\int \frac{d^3\vec{p}_{\nu_\beta}}{(2 \pi)^3} f(E_{\nu_\beta}, \mueff_{\nu_\beta}, T) \int \frac{d^3\vec{p}_{\bar \nu_\beta}}{(2 \pi)^3} f(E_{\bar \nu_\beta}, \mueff_{\bar \nu_\beta}, T) \sigma_{\nu_\beta \bar\nu_\beta \rightarrow Z' }|\vec{v_{\nu_\beta}}-\vec{v}_{\bar{\nu}_\beta}| \gamma_{Z'} v_{Z'} \frac{\hbar}{\Gamma_{Z'}^{tot}}}{\int \frac{d^3\vec{p}_{\nu_\beta}}{(2 \pi)^3} f(E_{\nu_\beta}, \mueff_{\nu_\beta}, T) \int \frac{d^3\vec{p}_{\bar \nu_\beta}}{(2 \pi)^3} f(E_{\bar \nu_\beta}, \mueff_{\bar \nu_\beta}, T) \sigma_{\nu_\beta \bar\nu_\beta \rightarrow Z' }|\vec{v_{\nu_\beta}}-\vec{v}_{\bar{\nu}_\beta}|},
  \label{eq:dlength}
\end{equation}
where $\sigma_{\nu_\beta \bar\nu_\beta \rightarrow Z' }$ is the cross section for neutrino-antineutrino coalescence. Since the temperature and, therefore, the energy distribution of neutrinos and antineutrinos producing $Z'$ depend on the distance from the center, $\ell_{Z'}$ also varies with this distance. We have computed $\ell_{Z'}$ for each shell described in Table~\ref{tab:shells}. 

For the range of $g_{\mu -\tau}$ such that $\sim 3 \, \textrm{m} \, <\ell_{Z'}< R_{ns}\sim 20$ km, we can assume that, during time span $\tau_{\nu \bar{\nu} \to Z’}$, neutrino-antineutrino pairs convert to $Z’$ and decay into $\nu \bar{\nu}$ at a distance $\ell_{Z'}$. That is, after a time interval of $\tau_{\nu \bar{\nu} \to Z’}$, $\nu \bar{\nu}$ take a step of $\ell_{Z'}$ inside the core in a random direction. After time $t$, they take $N=t/\tau_{\nu \bar{\nu} \to Z’}$ random steps, which takes them on average a distance $\sqrt{N} \ell_{Z'}$ far away from where they started. Thus, making the approximation that the whole proto-NS has the same temperature and density conditions, in order for a neutrino to travel a distance $R_{ns}$ (exit the star), 
$t=t_{\rm diff}^{\rm new}$ must satisfy
\begin{equation}
( t_{\rm diff}^{\rm new}/\tau_{\nu \bar{\nu} \to Z’})^{1/2} \ell_{Z'}= R_{ns}.  
\end{equation}
In other words, the neutrino diffusion time via the new mechanism will be 
\begin{equation}
t_{\rm diff}^{\rm new}= (R_{ns}/\ell_{Z'})^2 \tau_{\nu \bar{\nu}\to Z’}.   
\end{equation}

In order to be more realistic, we consider the different shells of Table~\ref{tab:shells}. Then, the shortening of the diffusion time due to the new interactions, in the limit in which $ 3 \, \textrm{m} \, <\ell_{Z'}< R_{ns}$, can be computed as
\begin{equation}
    t_{\rm diff}^{\rm new}= \sum_{k=1}^n \frac{\left( R_k^2-R_{k-1}^2\right)}{(\ell_{Z'}^{k})^2} \tau_{\nu \bar{\nu}\to Z’} ^k,
    \label{tdiffn}
\end{equation}
where $\tau_{\nu \bar{\nu}\to Z’}^k$ and $\ell_{Z'}^{k}$ are the time scale of the neutrino-antineutrino interaction and the $Z'$ decay length, respectively, in the $k-$shell. Note that for muon and tau neutrinos only the temperature changes between shells when computing these quantities.

It is important to remark here that previous cooling bounds, such as those of Ref.~\cite{Croon:2020lrf}, are based on the shortening of the neutrino flux duration by a half when the energy taken away by the new mediator is comparable with the one taken by neutrinos in the first $10$~s~\cite{Burrows:1988ah, Choi:1989hi, Raffelt:1990yz, Burrows:1990pk}, {\it i.e.}, $3\times 10^{53}$ erg. Here, we extend these \PRDnote{disfavoured} regions by arguing that the duration of the neutrino signal can be shortened even in the cases where the mediator decays inside (their trapping area). The total energy of the core in form of $\nu_\mu$, $\bar{\nu}_\mu$, $\nu_\tau$ and $\bar{\nu}_\tau$ at the onset of the cooling phase, computed using the input in Table \ref{tab:shells}, amounts to $\sim 3 \times 10^{51}$~erg. If $t_{\rm diff}^{\rm new}=0.1$~s, the energy transfer via the $Z'$ production will be comparable to the luminosity within the SM so the burst duration will be  reduced by half.

Let us now turn to the regime $\ell_{Z'}>R_{ns}$. 
In this case, each $\nu \bar{\nu}$ pair producing the $Z'$ on-shell will be transferred outside the neutrinosphere. The energy transfer rate per unit volume and time taken by the $Z'$ can be written as
\begin{equation}
    \mathcal{L}=\frac{1}{2\pi^2}\int d E_{\nu_{\beta}} f(E_{\nu_{\beta}}, \mueff_{\nu_\beta}, T ) E_{\nu_{\beta}}^{2} \int_{E^{min}_{\bar{\nu}_{\beta}}}^{\infty} 2 \frac{dE_{\bar{\nu}_{\beta}} }{32 \pi} f(E_{\bar{\nu}_{\beta}}, \mueff_{\bar{\nu}_\beta}, T) E_{\bar{\nu}_{\beta}} \left(\frac{m_{Z'}}{E_{\bar{\nu}_{\beta}} E_{\nu_{\beta}}} \right)^2  |\mathcal{\overline{M}}|_{ \nu_{\beta} \bar{\nu}_{\beta} \to Z' }^2, 
    \label{eq:lum1}
\end{equation}
where the factor $2$ at the beginning of the inner integral comes from considering both muon and tau neutrino flavours.

Thus, the total energy carried by the $Z'$ per unit time is
\begin{equation}
    L= \sum_{k=1}^n \frac{4 \pi}{3} (R_k^3 -R_{k-1}^3)\mathcal{L}_{k},
     \label{eq:lum2}
\end{equation}
and imposing $L \leq 3 \times 10^{52} \, \rm erg/s$ is equivalent to the upper bound obtained in Ref.~\cite{Croon:2020lrf} for the \lmultau model. Note that $\mathcal{L}_{k}$ corresponds to the $Z'$ energy transfer rate per unit volume and time in each shell, which depends on their temperature.

The following remarks are in order:
\begin{itemize}
\item
In the regime where $\ell_{Z'}>R_{ns}$, the flavor composition of neutrinos transferred by $Z'$ can be different from the SM prediction. In the \lmultau model, neutrinos reaching out of the neutrinosphere will be composed of $\nu_\mu$, $\nu_\tau$, $\bar{\nu}_\mu$ and $\bar{\nu}_\tau$ with equal spectra and vanishing $\nu_e$ and $\bar{\nu}_e$ components, up to the corrections from the contributions from the standard neutrino diffusion from the neutrinosphere. Of course, these neutrinos will then go through flavor conversion traversing the outer shells of the star (and if $\ell_{Z'}$ is larger than the whole star size including the envelope, in the vacuum between the star and the Earth). Fluxes of $\nu_e$ and $\bar{\nu}_e$ will be produced via oscillation but the ratio between electron neutrino and muon and tau neutrino fluxes, {\it i.e.}, $F_{\nu_e}/F_{\nu_{\mu, \tau}}$, as well as the equivalent for electron antineutrinos, $F_{\bar{\nu}_e}/F_{\mu, \tau}$, at Earth will be different from the SM prediction.

\item 
As pointed out in the literature before~\cite{Akita:2022etk,Fiorillo:2022cdq}, if the mediator can travel distances larger than $R_{ns}$ before decaying back to neutrinos, the energy spectrum of the neutrinos reaching the Earth will be harder than the SM prediction. This is a consequence of the transfer of neutrinos from the inner core directly outside without being thermalized in the outer core which has a lower temperature. As a result, in the regime $\ell_{Z'}>R_{ns}$, the measurement of the neutrino energy spectrum can reveal new physics, independently of the neutrino burst duration measurement. However, for $3~{\rm m}<\ell_{Z'}<R_{ns}$, the neutrinos from the hotter inner core will be transferred first to the outer layers where they can reach thermal equilibrium so they cool down before coming out of the proto-NS. For the parameter range that we rule out in this paper, the burst duration measurement provides a unique SN observable to test the model. 
\end{itemize}


\subsubsection{Implications for the parameter space of the \lmultau model}
\label{sub:models_smallg}
\begin{figure}[!t]
  \centering
  \includegraphics[width=.328\textwidth]{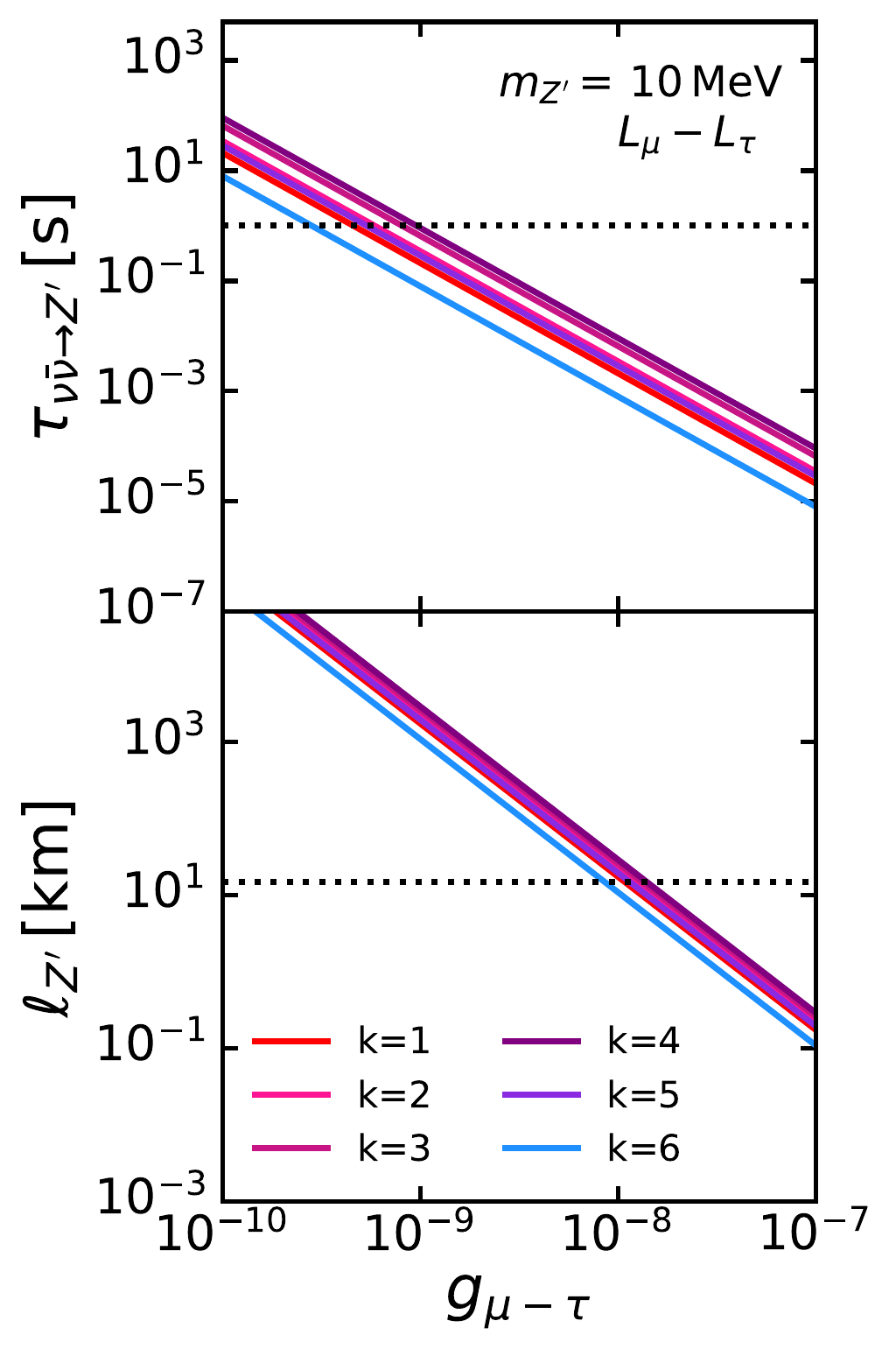}
  \includegraphics[width=.328\textwidth]{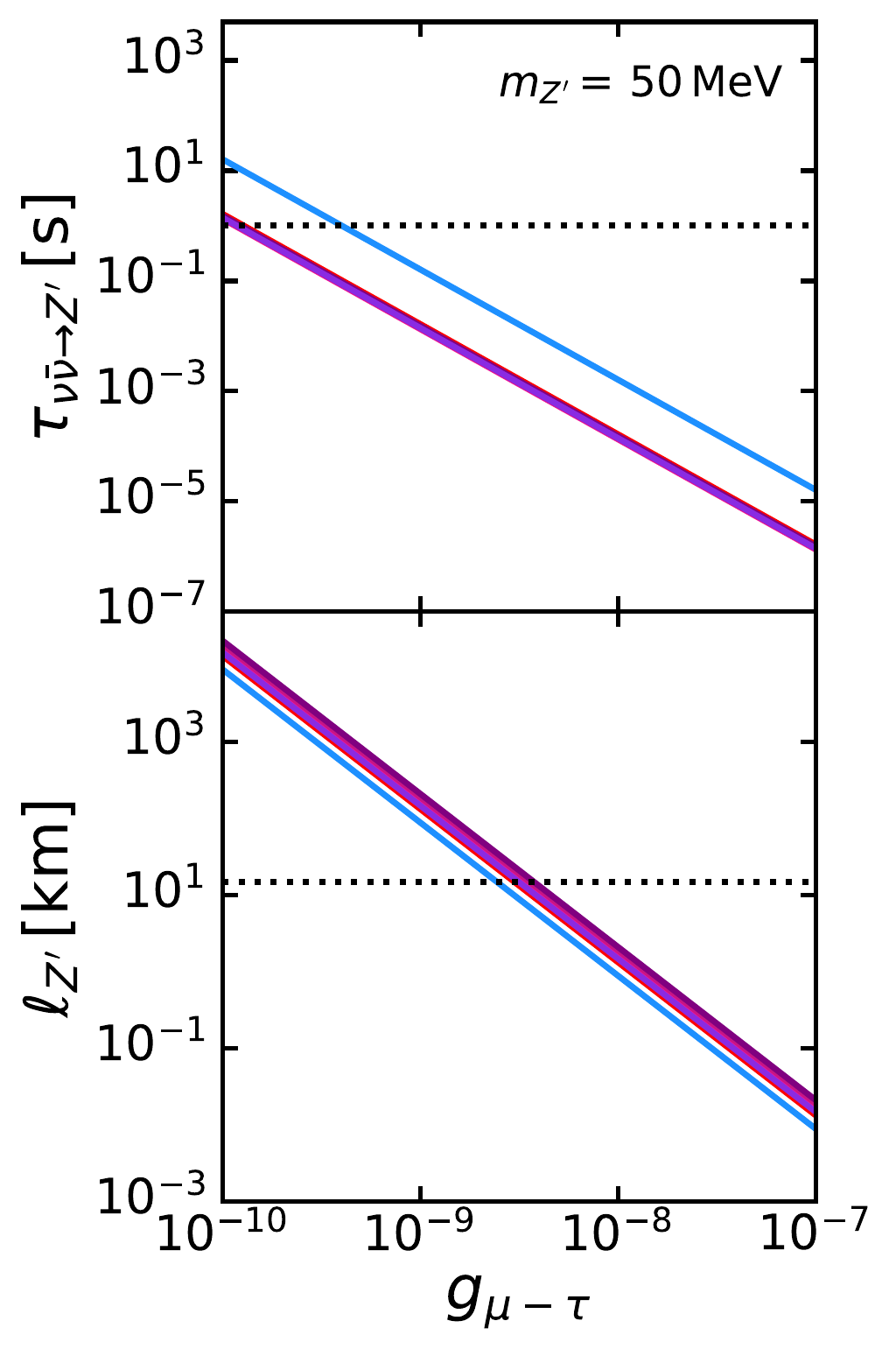}
  \includegraphics[width=.328\textwidth]{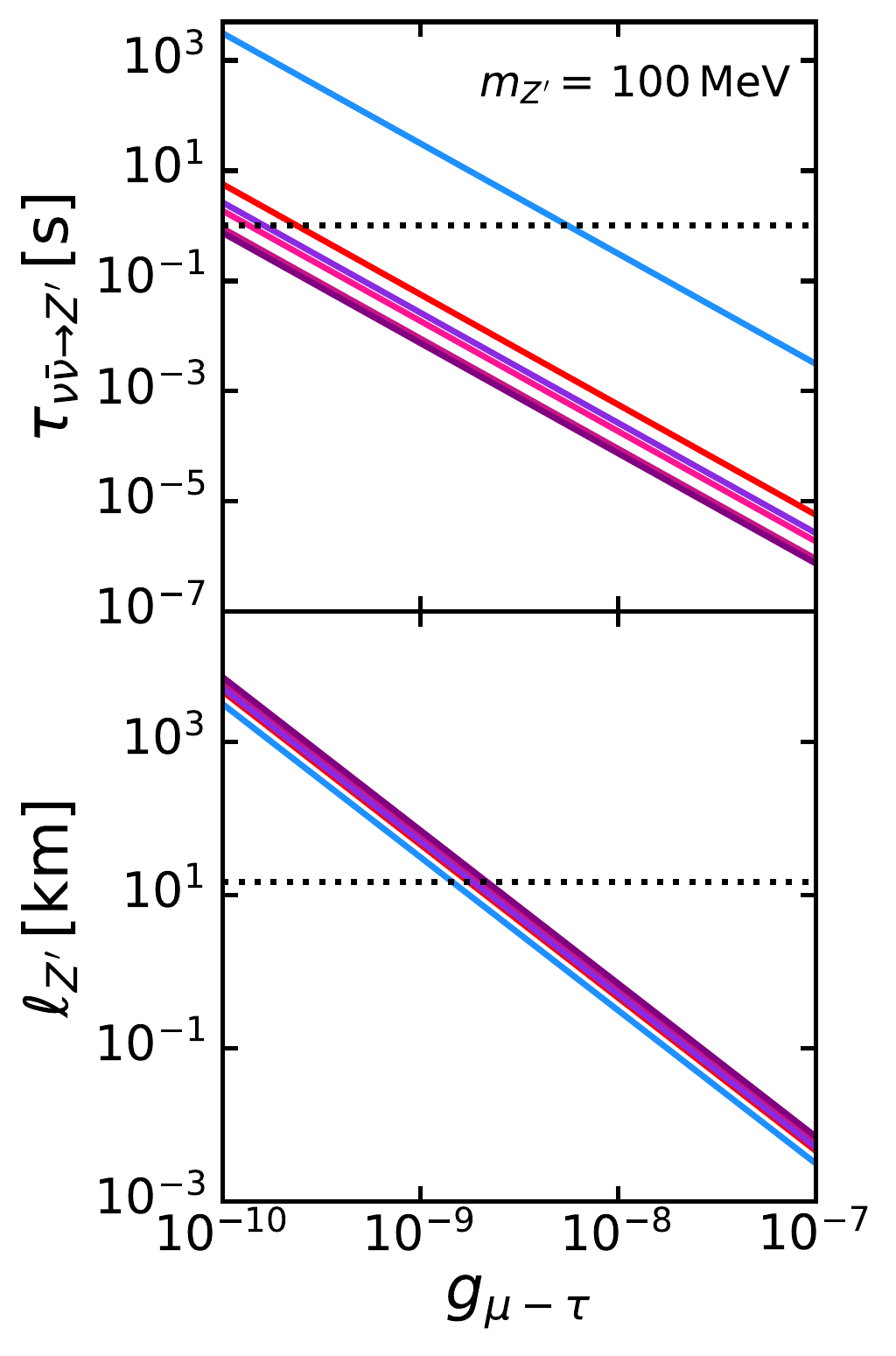}
  \caption{ The time scale of the neutrino-antineutrino interaction producing the $Z'$ vector mediator on-shell and the decay length of $Z'$ as a function of the coupling for the \lmultau model in the upper and lower panels, respectively. We show in different colours the corresponding values for each shell of the star. Different masses are depicted in the three columns showed. In particular, $m_{Z'}=10,\, 50,\, 100$ MeV for the left, central and right columns. For reference, the dotted lines correspond to the cases for which the timescale of the interaction is $1$ s, and where the decay length of the mediator is equal to the neutrinosphere radius. 
  }
  \label{fig:tscaledl_lmultau}
\end{figure}

\begin{figure}[!t]
  \centering
  \includegraphics[width=.6\textwidth]{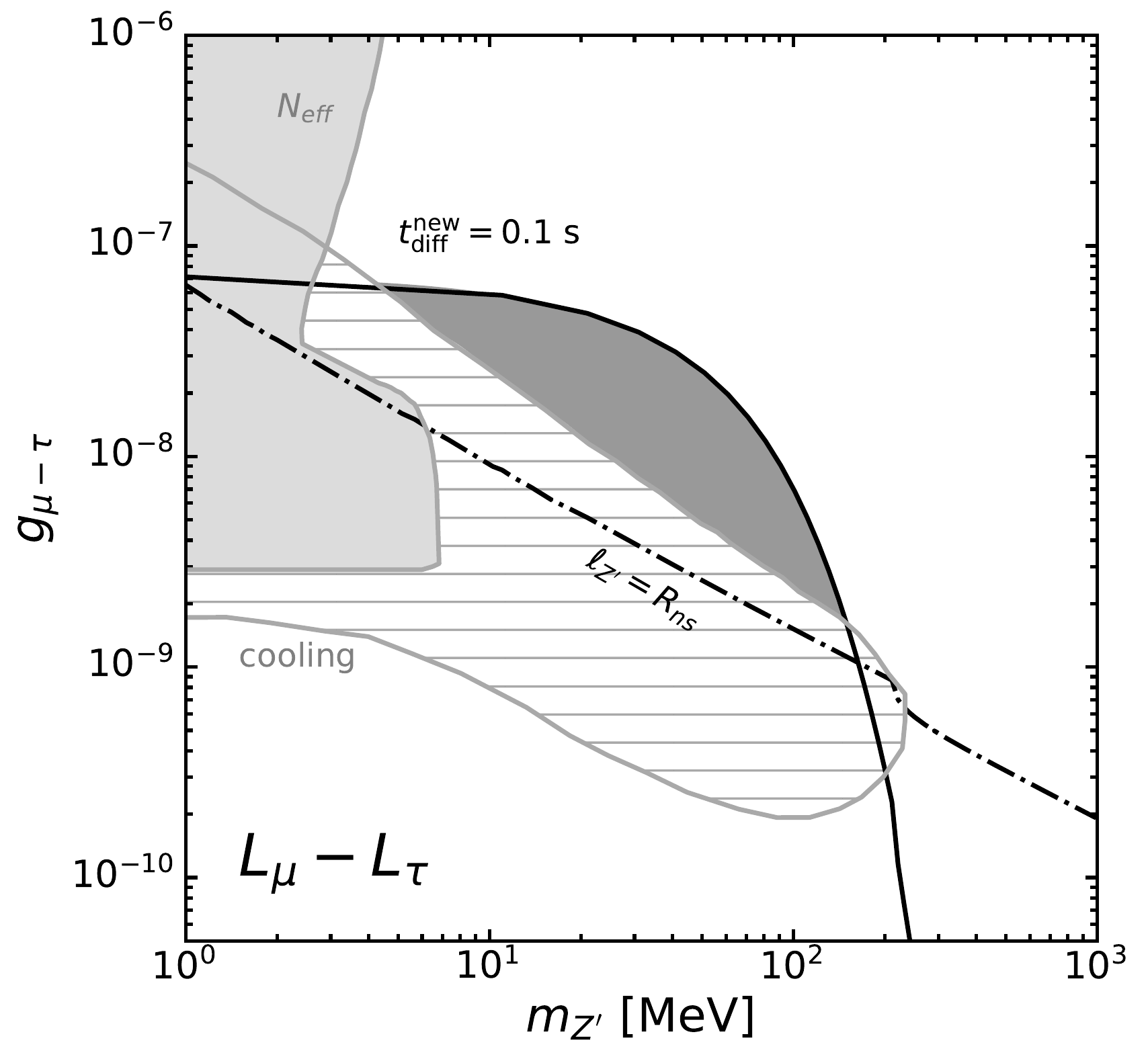}
  \caption{\PRDnote{Disfavoured} regions for the parameter space of the \lmultau model. The solid black line provides the lower bound on the coupling below which the neutrino duration signal would be reduced by more than half. The black dashed-dotted line indicates the region above which our calculation of the neutrino diffusion time holds. The dark grey region indicates the new excluded region by this work. In light grey we show the cosmological bounds from Big Bang nucleosynthesis \cite{Escudero:2019gzq} and in hatched grey the cooling bounds from Ref.~\cite{Croon:2020lrf}. 
  }
  \label{fig:bounds_lmultau_tdiff}
\end{figure}

Before discussing the \PRDnote{new disfavoured regions}, let us remind ourselves that all three neutrino flavors are in thermal equilibrium inside the core with each other as well as with the huge thermal energy source stored in matter fields (electrons and nucleons). In the \lmultau model, the $\nu_\mu\bar{\nu}_\mu$ and $\nu_\tau\bar{\nu}_\tau$ coalescence will not directly affect the $\nu_e$ population or its diffusion process. However, if there is an energy transfer mechanism such as the $\nu_\mu \bar{\nu}_\mu$ and/or $\nu_\tau \bar{\nu}_\tau$ mechanism depletes the energy content, once the initial $\nu_e$ leave the core via the standard diffusion mechanism with a time scale of $\sim 3$~s, there would be no energy left to replace them. Thus, the fact that $\nu_e$ are not directly affected does not maintain the SM prediction for the neutrino burst duration.

In the upper panels of Fig.~\ref{fig:tscaledl_lmultau}, we represent the time scale of $\nu_\mu \bar{\nu}_\mu \to Z'$ or $\nu_\tau \bar{\nu}_\tau \to Z'$ inside different shells of the proto-NS as a function of the gauge coupling, for various values of the mediator mass ($m_{Z'}=10,\, 50,\, 100$ MeV). The horizontal dotted line corresponds to $\tau_{\nu \bar\nu \to Z'}=1$~s. For values of $\tau_{\nu \bar\nu \to Z'}$ below this line, the coalescence rate of the neutrino-antineutrino pair is faster than their standard diffusion rate. The lower panels show $\ell_{Z'}$ for the different shells and the same choice of mediator masses, and the horizontal dotted line highlights the value of the neutrinosphere, $R_{ns}=20$~km. These two parameters, $\tau_{\nu \bar\nu \to Z'}$ and $\ell_{Z'}$, are key quantities for the calculation of the neutrino diffusion time due to new interactions, see \cref{tdiffn}.

Notice that $\ell_{Z'}$ is shorter in the outer shells because the $Z'$ produced with smaller temperatures have a smaller boost factor, $E_{Z'}/m_{Z'}$. Also, the $\nu\bar\nu\to Z'$ interaction rate is maximised when the temperature of the layer is similar to $\sim m_{Z'}/2\pi$ (since the average of neutrino and antineutrino energies for a given temperature of the star is $ \sim \pi T$) and consequently, $\tau_{\nu\bar\nu\to Z'}$ becomes smaller in those shells. This behaviour can be observed in the three columns of Fig.~\ref{fig:tscaledl_lmultau}. In particular, the value of $\tau_{\nu\bar\nu\to Z'}$ in the outer layer increases with $m_{Z'}$, since the temperature is lower and $Z'$ production is suppressed by the Boltzmann factor. For this reason, we do not include the last shell ($k=6$) when computing the neutrino diffusion time through the new interactions given by \cref{tdiffn}.

In~\cref{fig:bounds_lmultau_tdiff}, the dark grey area shows the new \PRDnote{disfavoured} region in the \lmultau parameter space obtained in this work. The solid black line provides the lower bound on the $g_{\mu-\tau}$ coupling below which the neutrino duration signal would be reduced by more than half. The black dashed-dotted line indicates the region above which our calculation of the neutrino diffusion time via the new interaction holds, {\it i.e.}, when the decay length of the mediator is smaller than the proto-NS radius. In light grey, we show the cosmological bounds from Big Bang nucleosynthesis based on the contribution to $N_{\rm eff}$ due to the presence of the mediator in the Early Universe~\cite{Escudero:2019gzq}. The cooling bounds from Ref.~\cite{Croon:2020lrf} can be observed in hatched grey. 
Using Eqs.~\eqref{eq:lum1} and \eqref{eq:lum2}, we have also confirmed that the lower limit from Ref.~\cite{Croon:2020lrf} is recovered for masses above $10$ MeV. We have found that the exact values of the bounds depend rather strongly on the assumptions on the temperature profile of the proto-NS.

 Our results extend the region of Ref.~\cite{Croon:2020lrf} \PRDnote{that is in tension with SN~1987A data}, and show that, even in the $\ell_{Z'}<R_{ns}$ regime, the neutrino diffusion time can be significantly shortened by the production of the $Z'$ via neutrino-antineutrino coalescence. In this regime, the $Z'$ decay length can be long enough to allow neutrinos to travel distances $\ell_{Z'}$ inside the star without significant interactions with nucleons, thus depleting the thermal energy stored in the proto-NS in form of muon and tau neutrinos and antineutrinos faster than in the SM scenario. In particular, for $m_{Z'}\sim 10-100$~MeV couplings below ${\cal O} (10^{-7})$ are \PRDnote{disfavoured}.

Notice that the treatment in Ref.~\cite{Croon:2020lrf} does not probe the parameter region marked with dark grey in Fig.~\ref{fig:bounds_lmultau_tdiff}. In this regime, the $Z'$ decays back inside the core (using the terminology of Ref.~\cite{Croon:2020lrf}, it is absorbed inside the core) and does not directly contribute to the luminosity. However, this overlooks the fact that when the produced $Z'$ bosons jump from the hot inner shells of the core to the outer shells, they warm up the outer shells, increasing the neutrino density and hence the energy emission from the outer shell. This effect is automatically ingrained in our treatment so we are able to probe the dark grey region. The dashed region excluded by Ref.~\cite{Croon:2020lrf} also includes a relatively narrow range above the dotted-dashed line at which $\ell_{Z'}< R_{ns}$. This is understandable because of two reasons: (1) The  $Z'$ decay is a stochastic process and a good fraction of $Z'$ decay long after $\ell_{Z'}/c$; (2) Due to simple geometry, the $Z'$ particles that are produced in the outer shells and head outwards travel less than $R_{ns}$ to reach outside the core. As a result, even above the dotted-dashed line, a good fraction of $Z'$ are not absorbed inside the core and directly contribute to the luminosity by decaying outside the neutrinosphere.


\PRDnote{One may also wonder whether the thermal distribution for neutrinos and antineutrinos is a good approximation to compute the production rate of $Z’$. In the relevant range of parameters shown in Fig.~\ref{fig:bounds_lmultau_tdiff} the time scale of processes that produce the $Z’$ is much larger than the time scale of weak interaction processes and, therefore, the local thermal equilibrium is expected to be maintained. This continues until the temperature throughout the core drops so much that the production of $Z’$ is not efficient anymore. According to our analysis this happens in less than $\sim 5$ s. Then the temperature reduces so much that the flux of neutrinos would be too low to be detectable by KamioKande and other detectors that observed SN 1987A. A similar argument has been widely invoked to obtain cooling bounds in the literature, such as for example in Refs.~\cite{Heurtier:2016otg, Croon:2020lrf, Shin:2021bvz}. However, in order to be conservative, due to the lack of a self-consistent simulation that includes the transport equations
for both the $Z'$ boson and neutrinos, we refer to the regions of the parameter space in tension with the SN~1987A measurements as ``disfavoured regions" instead of ``excluded" ones.}


It should be noted that the $Z'$ could potentially undergo interactions with neutrinos and muons before leaving the star. However, the rate of the scattering of $Z'$ off neutrinos, $Z'+\nu \to Z'+\nu$ is suppressed by $g_{\mu-\tau}^4$ and is therefore negligible. Despite the smallness of the muon density inside a proto-NS, the scattering $Z'+\mu \to \mu +\gamma$ can be dominant as it is only suppressed by $g_{\mu-\tau}^2$. Using the muon distribution derived in Ref.~\cite{Bollig:2020xdr}, we have computed the mean free path of $Z'+\mu \to \mu+\gamma$ and we have found it to be $\sim 60~{\rm km}\,(7\times 10^{-8}/g_{\mu-\tau})^2$. Thus, for $m_{Z'}>10$~MeV, the mean free path is larger than the decay length of $Z'$ by more than one order of magnitude. As a result, the $Z'$ decays long before undergoing any scattering.

Finally, the non-negligible amount of muons inside the proto-NS \cite{Bollig:2020xdr} could open a new production mechanism of $Z'$ via muon-photon semi-Compton scattering. However this production mechanism is only relevant for mediator masses below $10$ MeV \cite{Croon:2020lrf}, which are already ruled out by $N_{\rm eff}$ \cite{Bauer:2018onh} and cooling bounds \cite{Croon:2020lrf}. For masses above $10$ MeV the leading process producing the $Z'$ is neutrino-antineutrino coalescence and, therefore, neglecting the small amount of muons expected in the proto-NS is a good approximation for our parameter space range of interest.

\section{Conclusions}
\label{sec:conclusions}

In this article, we have investigated the on-shell production of low-mass vector mediators from neutrino-antineutrino interactions in the core of proto-NS, determining the conditions under which these processes significantly alter the neutrino signal duration in SN explosions. We point out that the on-shell production of a long-lived $Z'$ and its subsequent decay into neutrinos inside the proto-NS can significantly shorten the duration of the observed neutrino burst, providing a new way to test neutrino self-interactions.

As a concrete example, we have computed the rate of neutrino-antineutrino interactions in a well-motivated new physics scenario that features a new vector mediator, namely \lmultau. In order to calculate the rate of $\nu \bar{\nu}\to (Z')^*\to \nu \bar{\nu}$ in the nuclear medium, we have considered the radial dependence of the density, temperature and chemical potentials of the particles inside the proto-NS. We have found that, thanks to their vanishing chemical potentials, the rate of muon or tau neutrino-antineutrino interactions can exceed that of the SM scattering off nucleons when the kinematics allow for on-shell $Z'$ production, even for couplings as small as $g_{\mu-\tau}\sim 10^{-6}-10^{-5}$.

We have first discussed that, contrary to the previous claims, in the range of parameter space where neutrino-antineutrino interactions dominate over the SM neutrino-nucleon scattering, the diffusion time does not significantly change. This is because the momentum conservation preserves the outflow of neutrinos plus antineutrinos. Our argument confirms the results in Ref.~\cite{Dicus:1988jh}. In particular, the \lmultau solution for the muon magnetic moment anomaly is not ruled out for this reason. We have then examined whether frequent neutrino-antineutrino interactions can redistribute the energy of neutrinos and antineutrinos, leading to a change in the burst duration. However, we found out that the effects of energy redistribution are too small to lead to a discernible impact in the SN 1987A data. More work is needed to determine whether the subtle changes in the neutrino spectrum can be observable with future SN data.

It was shown by Ref.~\cite{Croon:2020lrf} that if the $Z'$ decays outside the star, the neutrino burst duration can be significantly reduced so that the measurement of the SN 1987A burst duration already rules out very small couplings of ${\cal O}(10^{-9})$ for the mass range of $10$~MeV$-200$~MeV. In this parameter space range, in addition to the neutrino burst duration, the energy spectrum as well as the flavor composition of the neutrino and antineutrino fluxes reaching the Earth will be significantly altered.

We have focused on the region of the parameter space for which the decay length of $Z'$ is larger than the standard neutrino mean free path, $\sim$~3 m, but smaller than the proto-NS radius. We show that, in this range, even if the $Z'$ decays inside the proto-NS, the neutrino burst duration can also be significantly reduced allowing us to \PRDnote{determine} new areas of the parameter space with couplings of $\sim 6 \times 10^{-8}$ \PRDnote{that are in tension with SN~1987A data}. We argue that this \PRDnote{criterion} is robust against uncertainties in the measurement of the time duration of the neutrino burst (unlike any bound based on SN 1987A neutrino burst duration measurement that one would impose on a model that predicts prolongation rather than shortening of the duration). In this new range of the parameter space, neutrinos from deep inner core are first transferred to the outer shells by the $Z'$ production and become thermalized in the cooler outer core. Thus, the energy spectra of neutrinos and antineutrinos coming out of the star are not expected to be significantly different from the standard prediction. Therefore, while the smaller couplings for which $\ell_{Z'}>R_{ns}$ can be alternatively tested by the energy spectrum measurement, in the range of parameters that ruled out for the first time in this paper, the neutrino burst duration holds a unique and special place of honor to explore new physics. Our results are summarized in Fig.~\ref{fig:bounds_lmultau_tdiff}.

These results are relevant for any other model that features new MeV scale mediators that couple to the neutrino sector such as light mediator models with NSI.

\section*{acknowledgements}

We would like to thank Juan Antonio Aguilar-Saavedra, Rafael Aoude, Miguel Escudero, Ivan Esteban, Patrick Foldenauer, Luca Mantani, and M. \'Angeles P\'erez-Garc\'ia, for useful discussions and comments. D. G. C. acknowledges support from the Comunidad de Madrid under Grant No. SI2/PBG/2020-00005. The work of M.C. was partially funded by the F.R.S.-FNRS through the MISU convention F.6001.19. Y. F. would like to acknowledge support from the ICTP through the Associates Programme and from the Simons Foundation through grant number 284558FY19 and from Saramadan under contract No.~ISEF/M/401439. She has also received funding/support from the European Union’s Horizon 2020 research and innovation programme under the Marie Skłodowska-Curie grant agreement No 860881-HIDDeN as well as under the Marie Skłodowska-Curie Staff Exchange  grant agreement No 101086085 – ASYMMETRY. This work is partially supported by the Spanish Agencia Estatal de Investigaci\'on through the grants PID2021-125331NB-I00 and CEX2020-001007-S, funded by MCIN/AEI/10.13039/501100011033.

\bibliographystyle{JHEP-cerdeno}
\bibliography{biblio}

\end{document}